\documentclass[final,3p,times]{elsarticle}

\usepackage{amsmath,hyperref}
\usepackage{lineno}

\usepackage{epstopdf}
\journal{Journal}

\usepackage{lineno,hyperref}
\usepackage{graphicx}
\usepackage{dcolumn}
\usepackage{bm}
\usepackage{epsfig}
\usepackage{booktabs}
\usepackage{subfigure}
\usepackage{graphics}
\usepackage{amssymb}
\usepackage{amsmath}
\usepackage{array}
\usepackage{color}
\usepackage{booktabs}
\usepackage{multirow}
\usepackage{caption}
\usepackage{chngpage}
\usepackage{subfigure}
\usepackage{mathrsfs,gensymb,float}

\biboptions{numbers,sort&compress}
\modulolinenumbers[1]
\begin{document}

\captionsetup[figure]{labelfont={bf},name={Fig.},labelsep=period}        

\begin{frontmatter}
	
	\title{Diffuse-interface modeling and simulation of the freezing of binary fluids with the Marangoni effect}


	\author[1]{Jiangxu Huang}
	\author[1,3,4,5]{Zhenhua Chai\corref{mycorrespondingauthor}}	
	\ead{hustczh@hust.edu.cn}
	\cortext[mycorrespondingauthor]{Corresponding author}
	\author[1]{Xi Liu}	
	\author[1]{Changsheng Huang}

	\address[1]{ School of Mathematics and Statistics, Huazhong University of Science and Technology, Wuhan 430074, China}
	\address[3]{Institute of Interdisciplinary Research for Mathematics and Applied Science,Huazhong University of Science and Technology, Wuhan 430074, China}	
	\address[4]{Hubei Key Laboratory of Engineering Modeling and Scientific Computing, Huazhong University of Science and Technology, Wuhan 430074, China}	
	
	\address[5]{The State Key Laboratory of Intelligent Manufacturing Equipment and Technology, Huazhong University of Science and Technology, Wuhan 430074, China}


\begin{abstract}
	
This paper proposes a diffuse-interface model for simulating gas-liquid-solid multiphase flows involving solid–liquid phase change, solute transport, and the Marangoni effect. In this model, a phase-field method is employed to capture the evolution of fluid–fluid interfaces, while an enthalpy-based approach is used to describe the temperature field and implicitly track the solid–liquid interface. Solute transport is modeled using a constrained scalar-transport model combined with a pseudo-potential concentration approach. The proposed diffuse-interface model satisfies the reduction consistency, and can degenerate to the conservative phase-field method for incompressible two-phase flow and the classical enthalpy method for binary material solidification in an appropriate way. Furthermore, the model not only can preserve the mass conservation, but also can capture the volume change induced by phase change. To solve the diffuse-interface model, a lattice Boltzmann (LB) method is then developed, and the numerical tests demonstrate that the method has a good performance in the study of the freezing process coupled with Marangoni flow, phase-change-induced volume change, and solute transport. Finally, the model is applied to investigate the freezing dynamics of a system containing an insoluble impurity, revealing the complex interaction between the advancing freezing front and the impurity. It is found that the numerical results are in good agreement with experimental data.

\end{abstract}

	
\end{frontmatter}

\section{Introduction}

Freezing or solidification, a common liquid-solid phase change phenomenon, is a ubiquitous phenomenon in both natural and industrial processes \cite{HuerreARFM2024,DuNRP2024}. In nature, freezing processes, from sea ice formation and permafrost evolution to dynamic glacier growth, have some significant influences on Earth's climate system and environmental changes \cite{HannaNature2013,PostScience2013,McGuirePNAS2018}. In industry, the solidification is an important process for some advanced technologies, including additive manufacturing \cite{ZhangNC2020}, thermal energy storage \cite{RostamiE2020}, spray freezing \cite{MohitDT2025}, and freeze casting \cite{WegstNRMP2024}. Therefore, a thorough understanding of the liquid-solid phase-change process is essential to explore the mechanisms governing these natural phenomena and to optimize the related industrial technologies.

As a fundamental problem, the freezing of a single pure droplet has been widely studied through theoretical and experimental approaches \cite{AndersonJCG1996,SnoeijerAJP2012,MarinPRL2014,ZhangATE2017,ZhangATE2019,TembelyJFM2019}. The formation of a sharp tip after the freezing of a droplet is an intriguing phenomenon that has prompted extensive studies on the mechanisms of tip singularity formation. Anderson et al. \cite{AndersonJCG1996} experimentally investigated the tip formation, developed a droplet freezing model for a droplet freezing on a plate through incorporating a dynamic triple-junction condition with contact line slippage to explain the formation of cusps during freezing. Subsequently, Snoeijer et al. \cite{SnoeijerAJP2012} reported that the tip formation arises from the solid-liquid density difference, and proposed an analytical geometric model to elucidate the quantitative relationship between the angle of the conical tip and the solid-liquid density difference. Due to the absence of a quantitative description of tip singularity, Marin et al. \cite{MarinPRL2014} conducted a systematic experimental measurement on the  conical tip angle and found that the cone angle is independent of surface temperature, contact angle, and freezing rate. In parallel with the studies on tip singularity, there are some works focused on predicting freezing rate and solidification time \cite{ZhangATE2017,ZhangATE2019,TembelyJFM2019}. For instance, Zhang et al. \cite{ZhangATE2017,ZhangATE2019} proposed a freezing model that accounted for the effect of supercooling and gravity. They also demonstrated that the assumption of a spherical freezing front, rather than a planar one, can be used to reduce the discrepancy in the prediction of freezing time from 10$\%$ to within 2$\%$ \cite{ZhangATE2017,ZhangATE2019}. Tembely et al. \cite{TembelyJFM2019} developed a one-dimensional heat-conduction model, and found that it not only can accurately predict the freezing time, but also can capture volume expansion and the tip singularity. In addition, there are some works focused on the effects of several external factors on the freezing process, including the substrate properties (e.g., surface structure and wettability) \cite{ChangATE2023}, gravity \cite{ZengPRF2022}, ambient humidity \cite{SebilleauPRE2021}, external energy field \cite{FangPRF2020}, and asymmetric cooling conditions \cite{StarostinJCIS2022}, and these studies have enriched the understanding of solidification of a single droplet.

As an alternative to the experimental and theoretical approaches, numerical methods have also been used to study the freezing processes. It is known that the freezing usually occurs in environmental fluids and is significantly affected by surrounding fluid dynamics, making it a multiphase flow problem involving solid-liquid phase change. However, due to the dynamic evolution of interfaces among different phases, and intricate coupling between fluid flow and heat transfer, it is still challenging to develop some numerical methods for such a complex problem. Actually, the numerical methods for modeling the freezing processes typically employ two interface-capturing approaches to track the solid-liquid and gas-liquid interfaces, and can be divided into three main categories, including the diffuse-interface method (e.g., phase field models \cite{HuangJCP2022,ZhangJCP2022,ZhangJCP2024,ZhangJCP2025,BrownSAE2023}, phase-field-enthalpy models \cite{MohammadipourJFM2024,HuangPRE2024,HuangIJHMT2025}, and pseudo-potential-enthalpy method \cite{ZhangPRE2020}), sharp-interface method (e.g., front-tracking method \cite{VuIJMF2015}, level-set method \cite{ShetabivashJCP2020}, VOF-IBM method \cite{LyuJCP2021} and level-set fluid-moment method \cite{YeCAMC2024}), and hybrid diffuse-sharp interface method (e.g., coupled level-set and enthalpy method \cite{ThirumalaisamyIJMF2023}, coupled phase-field and VOF method \cite{WeiJCP2025}). Although these numerical methods have achieved a great success in the simulation of the freezing of pure fluid droplets, there are many freezing processes that include binary or multi-component fluids (e.g., the water-solute system) \cite{LohseNRP2020}. Unlike pure fluids, the solutes are usually rejected at the solid-liquid interface during the binary freezing process, resulting in a significant concentration gradient. The concentration gradient would further bring an influence on the surface tension, driving the solutal Marangoni flow \cite{ScrivenNature1960}. On the other hand, the temperature gradient within the droplet during freezing also has a great influence on the surface tension, and triggers the thermal Marangoni flow. Under the solutal and thermal Marangoni effects, the flow field would become more complex, and some anomalous solute/temperature distributions are also observed during the freezing process \cite{WangPRL2024,LeiIJHMT2025}, which cannot be captured by some existing numerical methods that neglect both solutal and thermal Marangoni effects \cite{HuangJCP2022,ZhangJCP2022,ZhangJCP2024,ZhangJCP2025,BrownSAE2023,MohammadipourJFM2024,HuangPRE2024,HuangIJHMT2025,ZhangPRE2020,VuIJMF2015,ShetabivashJCP2020,LyuJCP2021,YeCAMC2024,ThirumalaisamyIJMF2023,WeiJCP2025}.

We note that although both solutal and thermal Marangoni effects have been widely considered in the numerical methods for welding \cite{TsaiIJNMF1989,LianCMAME2023}, crystal growth \cite{LanJCG2001} and evaporation \cite{DiddensJFM2021,MialheJCP2023}, existing numerical methods for the study of the freezing have failed to fully incorporate these effects and their potential coupling in the study of the freezing of binary droplet \cite{HuangJCP2022,ZhangJCP2022,ZhangJCP2024,ZhangJCP2025,BrownSAE2023,MohammadipourJFM2024,HuangPRE2024,HuangIJHMT2025,ZhangPRE2020,VuIJMF2015,ShetabivashJCP2020,LyuJCP2021,YeCAMC2024,ThirumalaisamyIJMF2023,WeiJCP2025}. To fill this gap, and to describe the Marangoni effects on the complex process of binary freezing, we first propose a diffuse-interface model for binary freezing by coupling the phase-field method and the enthalpy method. The model is developed based on three key aspects: (i) Following our recent diffuse-interface approach \cite{HuangPRE2024}, the phase-field method is used to capture the interface evolution between the ambient fluid and phase-change material, and the enthalpy method is adopted to implicitly track the solid-liquid interface and describe the temperature field. (ii) To account for the Marangoni effect on the flow field, the concept of continuum surface force is employed to model interfacial tension and Marangoni stresses. (iii) A constrained scalar-transport model combined with the pseudo-potential concentration approach is utilized to depict the solute transport in the multiphase flow systems within the current diffuse-interface framework. Furthermore, the present diffuse-interface model satisfies the reduction consistency property, which means that in the absence of phase change, it would reduce to the conservative phase-field model for incompressible two-phase flow \cite{WangCapillarity2019}, while in the absence of multiphase flow, it would degenerate into the classical enthalpy approach for binary phase-change problems \cite{ChakrabortyJFM2007}.

The rest of this paper is structured as follows. In Section \ref{sec2}, the physical problem of binary freezing and the diffuse interface model are presented, followed by the LB method developed in Section \ref{sec3}. In Section \ref{sec4}, a series of numerical experiments are performed to test the accuracy and efficiency of the LB method. Finally, some conclusions are given in Section \ref{sec5}.

\section{Problem statement and mathematical model}
\label{sec2}
\begin{figure}[H]
	\centering
	\includegraphics[scale=0.45]{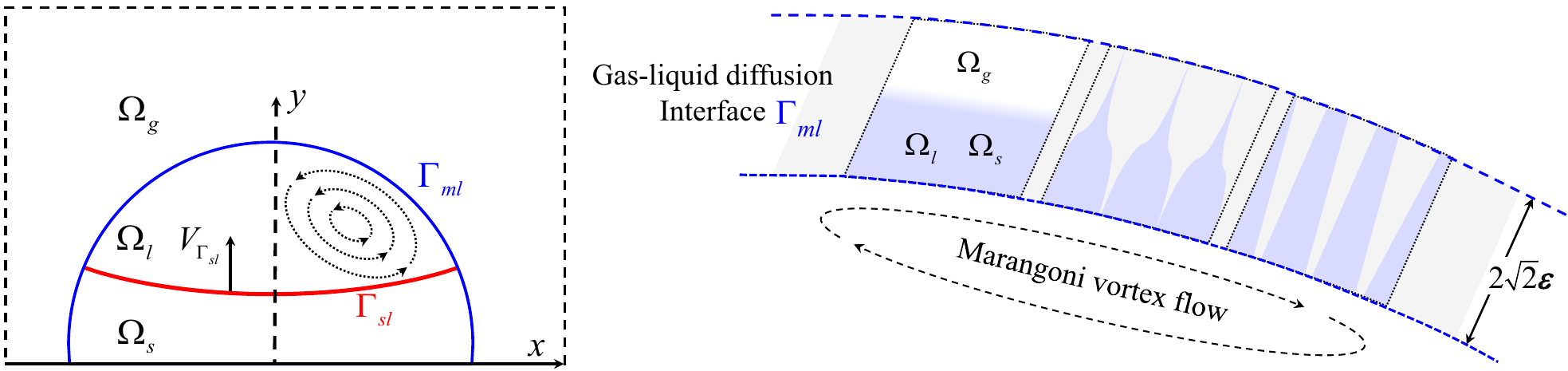}	
	\put(-440,93){(\textit{a})}
	\put(-250,93){(\textit{b})}	
	\caption{ Schematic diagram of the problem. (a) The freezing of a binary droplet on a cold substrate. The streamlines schematically illustrate the direction of the Marangoni flow inside the droplet. $\Omega_g$ ($\phi=-1 \cap f_s=0$), $\Omega_l$ ($\phi=1 \cap f_s=0$), and $\Omega_s$ ($\phi=1 \cap f_s=1$) are filled with gas, liquid and solid phases. $\Gamma_{sl}$ indicates the freezing front between the solid and liquid phases, and $\Gamma_{mg}$ represents the interface between the solid-liquid mixture and the gas phase. (b) A zoom-in map illustrates the interface between the solid-liquid mixture and the gas phase $\Gamma_{mg}$. From left to right, it displays the one-dimensional diffuse interface, the microstructure of this interface, and a close-up view of the microstructure, schematically indicating that the fingers are assumed to be thin structures.}
	\label{fig1}
\end{figure}

In this section, we will present a diffuse-interface model for the freezing of a binary fluid in the presence of gas phase, and mainly focus on the quasi-equilibrium isotropic solidification process without the formation of protrusions or dendritic crystal structures. We consider the classical problem of a binary droplet freezing on a cold wall as a representative example. As shown in Fig. \ref{fig1}, in the computational domain $\Omega = \Omega_l \cup \Gamma_{sl} \cup \Omega_s \cup \Gamma_{mg} \cup \Omega_g$, $\Gamma_{sl}$ represents the solid-liquid phase-change interface or freezing front, and $\Gamma_{mg}$ denotes the interface between the solid-liquid mixture and the gas phase. The domains $\Omega_g$, $\Omega_l$, and $\Omega_s$ are filled with the gas, liquid, and solid phases which are labeled by the subscripts $g$, $l$, and $s$. When the liquid phase temperature decreases below the freezing temperature $T_m$, the liquid changes into the solid phase, and simultaneously, the solid-liquid interface $\Gamma_{sl}$ advances. During the phase-change process, the volume expansion or shrinkage caused by the solid-liquid density difference also leads to the movement of $\Gamma_{mg}$. To capture the moving interfaces, an enthalpy method is used to implicitly track the freezing front $\Gamma_{sl}$ , while a phase-field method is adopted to capture the gas–liquid interface $\Gamma_{mg}$. Within this framework, the distinct phases are identified by using a phase-field order parameter $\phi$ and a solid fraction $f_s$. In addition, a key feature of binary fluid solidification is the migration and redistribution of solute due to different solubilities in different phases, which would induce a concentration gradient along the fluid interface. Both the concentration gradient and temperature gradient further bring a significant influence on surface tension, which in turn drive the solutal and thermal Marangoni flows, respectively. Before introducing the mathematical model for this complex system, the following principles and assumptions are clarified:
\begin{itemize}
	\item The fluid is assumed to be Newtonian and immiscible, and the multiphase flow is in a laminar state. 
	\item The gas phase is not directly involved in the phase change process, but it acts as a moving and deformable boundary in the solidification. 
	\item The volume change during the freezing is induced only by the density difference between the solid and liquid phases.
	\item The fluid flow is restricted to the gas-liquid two-phase region, while the distributions of temperature and concentration are continuous across the entire domain.
	\item The physical properties of fluids are independent of the temperature and concentration of solute except for surface tension.
\end{itemize}
In what follows, we will present a diffuse-interface model for binary freezing in the presence of a gas phase, which is composed of the phase-field equation for the gas-liquid interface, the Navier-Stokes equations for the flow field, the enthalpy-based energy equation for the temperature field, and the advection-diffusion equation for the concentration field.

The phase-field equation, as one of the most popular methods, has been widely used to describe moving interface problems through introducing an order parameter. A distinct feature of this model is that it does not need to explicitly track the moving interface, and has the ability to handle the complex topological change of the interface. Actually, there are two main types of phase-field equations, i.e., the Cahn-Hilliard equation \cite{CahnJCP1958} and the Allen-Cahn equation \cite{AllenAM1976}. Compared to the fourth-order Cahn-Hilliard equation, the second-order Allen-Cahn equation is much simpler to discretize and solve numerically, resulting in a lower computational cost \cite{WangCapillarity2019}. For this reason, in this work we will consider the second-order conservative Allen-Cahn equation for capturing the gas–liquid interface. Following the work of Sun and Beckermann \cite{SunJCP007}, the advection equation for the order parameter can be expressed as,
\begin{equation}
	\frac{\partial \phi}{\partial t}+\left(u_n \mathbf{n}+\mathbf{u}\right) \cdot \nabla \phi=0,
	\label{eq1}
\end{equation}
where $\phi$ is the phase-field order parameter, $\mathbf{u}$ is the fluid velocity, $\mathbf{n}=\nabla \phi / |\nabla \phi|$ is the unit vector normal to the interface. $u_n$ is the normal interface speed, and is defined as $u_n=-M_\phi \kappa_\phi$. Here $M_{\phi}$ is the mobility, $k_{\phi}$ is the interface curvature, and is defined as
\begin{equation}
	\kappa=\nabla \cdot \mathbf{n}=\nabla \cdot\left(\frac{\nabla \phi}{|\nabla \phi|}\right)=\frac{1}{|\nabla \phi|}\left[\nabla^2 \phi-\frac{(\nabla \phi \cdot \nabla)|\nabla \phi|}{|\nabla \phi|}\right].
	\label{eq2}
\end{equation}
For the simple one-dimensional problem, the equilibrium distribution of the order parameter can be approximated by the following hyperbolic tangent profile,
\begin{equation}
	\phi(x)=\frac{\phi_1+\phi_2}{2}+\frac{\phi_1-\phi_2}{2} \tanh \left(\frac{x}{\sqrt{2}\epsilon}\right),
	\label{eq3}
\end{equation}
where $\epsilon$ is a measure of the interface thickness, $\phi_1$ and $\phi_2$ are two constants corresponding to phases 1 and 2. With the help of Eq. (\ref{eq3}), one can determine the gradient of $\phi$ and its normal as
\begin{equation}
	|\nabla \phi|=\frac{d \phi}{d x}=\sqrt{\frac{2 \beta}{k}}\left(\phi_1-\phi\right)\left(\phi-\phi_2\right), 		\frac{(\nabla \phi \cdot \nabla)|\nabla \phi|}{|\nabla \phi|}=\frac{4 \beta}{k}\left(\phi-\phi_1\right)\left(\phi-\phi_2\right)\left(\phi-\frac{\phi_1+\phi_2}{2}\right),
	\label{eq4}
\end{equation}
where $k$ and $\beta$ are two positive constant. In order to remove the fluid motion induced by curvature \cite{SunJCP007, ChiuJCP2011}, the counter term approach introduced by Folch et al. \cite{FolchPRE1999} is adopted, and with the help of Eqs. (\ref{eq2}) and (\ref{eq4}), one can rewrite Eq. (\ref{eq1}) as
\begin{equation}
	\begin{aligned}
		\phi_t+\mathbf{u} \cdot \nabla \phi  =M_\phi\left[\nabla^2 \phi-\frac{4 \beta}{k}\left(\phi-\phi_1\right)\left(\phi-\phi_2\right)\left(\phi-\frac{\phi_1+\phi_2}{2}\right)\right] -M_\phi\left[|\nabla \phi| \nabla \cdot\left(\frac{\nabla \phi}{|\nabla \phi|}\right)\right].
	\end{aligned}
	\label{eq5}
\end{equation}
Then following the procedure in Ref. \cite{ChiuJCP2011}, we can reformulate Eq. (\ref{eq5}) into a conservative form,
\begin{equation}
	\begin{aligned}
		\phi_t+\mathbf{u} \cdot \nabla \phi = M_\phi\left[\nabla^2 \phi-\nabla \cdot\left(\sqrt{\frac{2 \beta}{k}}\left(\phi_1-\phi\right)\left(\phi-\phi_2\right) \frac{\nabla \phi}{|\nabla \phi|}\right)\right],
	\end{aligned}
	\label{eq6}
\end{equation}
which is considered as the local AC equation \cite{GeierPRE2015}.

In this work, we adopt $\phi_1 = 1$ to denote the fluid undergoing phase change, while the surrounding fluid without phase change is labeled by $\phi_2 = -1$. In this case, the interface between the two fluids is given by $\Gamma={x|\phi(x)=0}$. Additionally, we set $k = 1$ and $\beta = 1/(4\epsilon^2)$, and Eq. (\ref{eq6}) can be written as \cite{ChaiIJHMT2018}
\begin{equation}
	\frac{\partial \phi}{\partial t}+\mathbf{u} \cdot \nabla \phi=\nabla \cdot M_\phi\left(\nabla \phi-\frac{\nabla \phi}{|\nabla \phi|} \frac{1-\phi^2}{\sqrt{2} \epsilon}\right).
	\label{ACE}
\end{equation}

Apart from the above Allen-Cahn equation for capturing the interface, we now focus on the governing equations for the fluid flows. To overcome the difficulty in directly treating the no-slip boundary condition imposed on the solid-fluid interface $\Gamma_{sl}$, several sharp-interface numerical methods have been developed, such as the immersed boundary method or the embedded boundary method \cite{LyuJCP2021}. However, these numerical approaches are prone to encountering an instability issue, especially for the mushy region formed in the freezing process \cite{WeiJCP2025}. On the contrary, the diffuse-interface method seems more robust for modeling fluid movement within such a mushy region. Particularly, the diffuse-interface method can enforce the velocity field to vanish in the solid region by introducing a penalty term into the momentum equation. In this work, we consider the following Navier-Stokes equations for the fluid flows \cite{HuangPRE2024}, 
\begin{equation}
	\nabla \cdot \mathbf{u}=\dot{m}, \dot{m}=\left(1-\frac{\rho_s}{\rho_l}\right) \frac{\partial f_s}{\partial t},
	\label{NS1}
\end{equation}
\begin{equation}
	\frac{\partial \rho \mathbf{u}}{\partial t}+\nabla \cdot(\rho \mathbf{u u})=-\nabla p+\nabla \cdot\left[\mu\left(\nabla \mathbf{u}+(\nabla \mathbf{u})^{\mathrm{T}}\right)\right]+\mathbf{F}_s+\rho \mathbf{f}+\mathbf{F}_b,
	\label{NS2}	
\end{equation}
which are formulated in the diffuse-interface framework. Here $\rho$ is the density, $\mu$ is the dynamic viscosity, $p$ is the pressure, $\mathbf{F}_b$ is the body force. $ \mathbf{f} $ is the force generated by fluid-solid interaction to be determined below, and can be used to enforce the no-slip boundary condition on the fluid-solid interface. The source term $\dot{m}$ is used to characterize the volume expansion or shrinkage induced by density change during the solid-liquid phase change process, and can be derived from the mass conservation of the phase change material \cite{HuangPRE2024}. For the surface tension force $\mathbf{F}_s$, Kim \cite{KimJCP2005}  proposed a general continuous form, which was further extended to Cartesian coordinates by Liu et al. \cite{LiuPRE2013},
\begin{equation}
	\mathbf{F}_s=\left(-\sigma \kappa \mathbf{n}+\nabla_s \sigma\right) \delta,
\end{equation}
where $\kappa=\nabla \cdot \mathbf{n}$ is the local interface curvature, $\nabla_s=(\mathbf{I}-\mathbf{n n}) \cdot \nabla$ is the surface gradient operator, and $\delta$ is the Dirac delta function, which can be expressed as $\delta=\alpha|\nabla \phi|^2$ and satisfies the following relation, 
\begin{equation}
	\int_{-\infty}^{+\infty} \delta d x=\frac{2 \beta \alpha}{k} \sqrt{\frac{k}{2 \beta}} \int_{\phi_2}^{\phi_1}\left(\phi_1-\phi\right)\left(\phi-\phi_2\right) d \phi=1.
\end{equation}
When the phase-field variable is varied from -1 to 1, we can determine $\alpha=3 \sqrt{2} \epsilon/4$. In such a case, the interfacial force can be rewritten  as

\begin{equation}
	\begin{aligned}
		\mathbf{F}_s & =\alpha \nabla \cdot\left[\sigma|\nabla \phi|^2 \mathbf{I}-\sigma \nabla \phi \nabla \phi\right] \\
		& =\alpha\left[|\nabla \phi|^2 \nabla \sigma+\sigma \nabla|\nabla \phi|^2-\nabla \sigma \cdot(\nabla \phi \nabla \phi)-\sigma \nabla \cdot(\nabla \phi \nabla \phi)\right].
	\end{aligned}
\end{equation}
Under the relation of $\nabla \cdot(\nabla \phi \nabla \phi)=\nabla\left(|\nabla \phi|^2/2\right)+\nabla \phi \nabla^2 \phi$, the interfacial force can be simplified as
\begin{equation}
	\begin{aligned}
		\mathbf{F}_s &=\frac{3 \sqrt{2} \varepsilon}{4}\left[|\nabla \boldsymbol{\phi}|^2 \nabla \sigma+\frac{1}{2} \sigma \nabla|\nabla \boldsymbol{\phi}|^2-\nabla \sigma \cdot(\nabla \boldsymbol{\phi} \nabla \boldsymbol{\phi})-\sigma \nabla^2 \boldsymbol{\phi} \nabla \boldsymbol{\phi}\right] \\
		& =\underbrace{\frac{3 \sqrt{2} \varepsilon}{4}\left[|\nabla \boldsymbol{\phi}|^2 \nabla \sigma-\nabla \sigma \cdot(\nabla \phi \nabla \boldsymbol{\phi})\right]}_{\text {Marangoni force }\left(\mathbf{F}_{\mathbf{s}, \mathbf{M}}\right)}+\underbrace{\frac{3 \sqrt{2} \varepsilon}{4} \frac{\sigma}{\varepsilon^2} \mu_\phi \nabla \boldsymbol{\phi}}_{\text {Capillary force }\left(\mathbf{F}_{\mathbf{s}, \mathbf{C}}\right)},
	\end{aligned}
	\label{Fs}
\end{equation}
where $\mu_\phi$ is the chemical potential, and can be given as \cite{WangCapillarity2019}
\begin{equation}
	\mu_\phi=4 \beta\left(\phi-\phi_1\right)\left(\phi-\phi_2\right)\left(\phi-\frac{\phi_1+\phi_2}{2}\right)-k \nabla^2 \phi.
\end{equation}
In addition, the surface tension coefficient is usually a function of temperature and solution concentration, and the relation can be described by a state equation. The state equation may be linear or nonlinear, depending on the physical properties of the system, while for simplicity, we only consider the following linear equation of state\cite{LiuPRE2013, ZhanPRE2010}, 
\begin{equation}
	\sigma(T,c)=\sigma_0+\sigma_T\left(T-T_0\right)+\sigma_c\left(c-c_0\right),
\end{equation}
where $\sigma_0=\sigma(T_0,c_0)$ is the surface tension coefficient at the reference temperature $T_0$ and concentration $c_0$, $\sigma_T=\partial \sigma / \partial T$ and $\sigma_c=\partial \sigma / \partial c$ are the rates of change of surface tension coefficient with temperature and concentration, respectively.

Besides the above hydrodynamic and phase-field equations, the energy equation based on enthalpy is adopted to describe the solid-liquid phase change process, in which the enthalpy can also be used to capture the evolution of the solid-liquid interface, and can be expressed as \cite{ChakrabortyJFM2007, HuangIJHMT2013}
\begin{equation}
	\frac{\partial(\rho H)}{\partial t}+\nabla \cdot\left(\rho C_p T \mathbf{u}\right)=\nabla \cdot(k \nabla T)+\rho C_p T \dot{m},
	\label{Tem}
\end{equation}
where $C_p$ is the specific heat capacity, $T$ is the temperature, $\lambda$ is the thermal conductivity, $H=C_p T+L f_l $ is the mixture total enthalpy with $L$ being the latent heat. The last term on the right-hand side of the above equation is derived from $\nabla \cdot \mathbf{u}=\dot{m}$, which can be neglected for the incompressible fluid flow. Then, the temperature $T$ and the liquid phase fraction $f_l$ can be uniquely determined by the mixture total enthalpy $\rho H$ \cite{ZhaoAML2020},
\begin{equation}
	f_l =\left\{\begin{array}{ll}
		0 & \rho H< \rho_s H_s \\
		\frac{\rho H - \rho_s H_s}{\rho_l H_l-\rho_s H_s} &\rho_s  H_s \leqslant \rho H \leqslant \rho_l H_l, \\
		1 & \rho H>\rho_l H_l
	\end{array} \quad T= \begin{cases} \frac{\rho H}{\rho_s  C_{p,s}}  & \rho H< \rho_s H_s \\
		T_s+\frac{\rho H-\rho_s H_s}{\rho_l H_l-\rho_s H_s}\left(T_l-T_s\right) & \rho_s H_s \leqslant \rho H \leqslant \rho_l H_l, \\
		T_l+ \frac{\rho H-\rho_l H_l}{\rho_l C_{p,l}}  & \rho H>\rho_l H_l\end{cases}\right.
	\label{Hfl}
\end{equation}
where $T_s$ and $T_l$ are the solidus and liquidus temperatures, respectively. $H_s$ and $H_l$ are the mixture total enthalpies corresponding to the solidus and liquidus temperatures, respectively. It is worth noting that the present formulas for $f_l$ and $T$ take into account the density difference of the solid and liquid phases, while it has been ignored in the previous work \cite{HuangPRE2024}, leading to an inconsistency with the flow field \cite{ZhaoAML2020}. In addition, the solid–liquid interface can be implicitly tracked in the computational domain by the solid fraction variable $f_s=1-f_l$ that is defined over the entire domain. $f_s$ is defined as 0 in the liquid phase, $f_s=1$ in the solid phase, and $0<f_s<1$ in the mushy zone. 

To describe the evolution of solute concentration in a binary fluids system undergoing solid–liquid phase change, we first present the model for solute transport in a two-phase system without phase change, which consists of the transport equations in the bulk regions of the two phases and the boundary conditions at the interface. For the bulk region of phase $i$, the transport equation for the solute concentration $\tilde{c}_i$ is given by \cite{MirjaliliIJHMT2022,ChenPRE2024},
\begin{equation}
	\frac{\partial \widetilde{c}_i}{\partial t}+\frac{\partial\left(\widetilde{c}_i \mathbf{u}\right)}{\partial x}=\frac{\partial}{\partial x}\left(D_i \frac{\partial \widetilde{c}_i}{\partial x}\right), \quad i=1,2 .
	\label{c1}
\end{equation}
where the subscript $i$ denotes phase index with $i=1$ and $i=2$ corresponding to the liquid and gas phases. $\widetilde{c}_i$ is the local solute concentration per unit volume of phase $i$, and $D_i$ is the diffusion coefficient. At the interface, the continuity of normal fluxes should be satisfied,
\begin{equation}
	D_1 \nabla_n \widetilde{c_1}=D_2 \nabla_n \widetilde{c_2},
	\label{c2}
\end{equation}
where $\nabla_n$ denotes the normal derivative with respect to the interface. It should be noted that the second-order partial differential equation (\ref{c1}) requires an additional boundary condition, which is provided by the chemical equilibrium at the interface, i.e., the Henry’s law,
\begin{equation}
	\widetilde{c_1}/\widetilde{c_1}=H,
	\label{c3}
\end{equation}
where $H$ is the constant Henry coefficient characterizing solute solubility across the interface.

We note that the model consisting of Eqs. (\ref{c1})–(\ref{c3}) is developed in the sharp-interface framework where the physical variables and parameters across the interface are discontinuous, while in the present diffuse-interface framework, they change continuously across the interface through the phase-field variable $\phi$, which varies smoothly from 1 (liquid phase 1) to -1 (gas phase 2). As a result, the aforementioned sharp-interface model should be reformulated. To this end, Meijer et al. \cite{MirjaliliIJHMT2022} introduce a hypothetical microscopic structure, in which the diffuse interface is represented as an array of finger-like structures between different phases. By applying the sharp-interface equations~(\ref{c1})–(\ref{c3}) for a two-dimensional system and performing volume averaging along the $y$ direction yields,
\begin{equation}
	\frac{\partial \overline{\widetilde{c_i}}}{\partial t}+\frac{\partial\left(\overline{\widetilde{c_i}} \mathbf{u}\right)}{\partial x}=\frac{\partial}{\partial x}\left(D_i \frac{\partial \overline{\widetilde{c_i}}}{\partial x}\right),
	\label{c4}
\end{equation}
where $\overline{\widetilde{c}_i}$ denotes the $y$-averaged concentration in phase $i$, and they are related to the conserved concentration per total volume $c_i$ through $\overline{\widetilde{c_1}}(x)=2 c_1(x) /[1+\phi(x)]$ and $\overline{\widetilde{c_2}}(x)=2 c_2(x) /[1-\phi(x)]$, here $c_i$ is the amount of scalar in phase $i$ per total volume. Then, one can obtain the governing equations for $c_i(x)$,
\begin{subequations}
	\begin{equation}
		\frac{\partial c_1}{\partial t}+\frac{\partial\left(c_1 \mathbf{u}\right)}{\partial x}=\frac{\phi+1}{2} \frac{\partial}{\partial x}\left(D_1 \frac{\partial \tilde{c}_1}{\partial x}\right),
		\label{c5a}
	\end{equation}
	\begin{equation}
		\frac{\partial c_2}{\partial t}+\frac{\partial\left(c_2 \mathbf{u}\right)}{\partial x}=\frac{1-\phi}{2} \frac{\partial}{\partial x}\left(D_2 \frac{\partial \tilde{c}_2}{\partial x}\right).
		\label{c5b}
	\end{equation}	
	\label{c5}
\end{subequations}
In the present phase-field model, the phase equilibrium at the diffuse interface results in a hyperbolic tangent distribution of the phase-field variable $\phi$, satisfying
\begin{equation}
	\nabla \phi=\frac{1-\phi^2}{\sqrt{2} \epsilon} \frac{\nabla \phi}{|\nabla \phi|}.
\end{equation}
With the help of the above equation, Eq. (\ref{c5}) can be rewritten as
\begin{subequations}
	\begin{equation}
		\frac{\partial c_1}{\partial t}+\nabla \cdot\left(c_1 \mathbf{u}\right)=\nabla \cdot\left[D_1\left(\nabla c_1-\frac{(1-\phi) c_1}{\sqrt{2} \epsilon} \mathbf{n}\right)\right],
		\label{c6a}
	\end{equation}
	\begin{equation}
		\frac{\partial c_2}{\partial t}+\nabla \cdot\left(c_2 \mathbf{u}\right)=\nabla \cdot\left[D_2\left(\nabla c_2+\frac{(1+\phi) c_2}{\sqrt{2} \epsilon} \mathbf{n}\right)\right],
		\label{c6b}		
	\end{equation}	
\end{subequations}
which are known as the two-scalar solute transport model. In general, the scalar can be transferred between the two phases, and an additional source term needs to be added \cite{MirjaliliIJHMT2022}. However, the solute diffusivity in phase 1 is a finite value $D$, while it is about zero in phase 2. Therefore, the concentration transport equation in the entire domain without solid-liquid phase change is given by
\begin{equation}
	\frac{\partial c}{\partial t}+\nabla \cdot(c \mathbf{u})=\nabla \cdot\left[D\left(\nabla c-\frac{(1-\phi) c}{\sqrt{2} \epsilon} \mathbf{n}\right)\right],
	\label{c7}	
\end{equation}
where $c$ is the solute concentration, and the flux term $\mathbf{J} = -\frac{(1 - \phi)c}{\sqrt{2}\epsilon} \mathbf{n}$ prevents solute diffusion into phase 2. Next, we consider the solute transport in the solid-liquid phase change process, and based on Eq. (\ref{c7}), one can derive the following governing equations,
\begin{subequations}
	\begin{equation}
		\frac{\partial c_s}{\partial t}+\nabla \cdot\left(c_s \mathbf{u}\right)=\nabla \cdot\left[D_s\left(\nabla c_s-\frac{(1-\phi) c_s}{\sqrt{2} \epsilon} \mathbf{n}\right)\right],
		\label{cs1}
	\end{equation}
	\begin{equation}
		\frac{\partial c_l}{\partial t}+\nabla \cdot\left(c_l \mathbf{u}\right)=\nabla \cdot\left[D_l\left(\nabla c_l-\frac{(1-\phi) c_l}{\sqrt{2} \epsilon} \mathbf{n}\right)\right].
		\label{cs2}
	\end{equation}	
	\label{cs}
\end{subequations}
At the solid–liquid interface $\Gamma_{sl}$, solute conservation yields
\begin{equation}
	\left(c_{l, \Gamma_{sl}}-c_{s, \Gamma_{sl}}\right) V_{\Gamma_{sl}}=[(D \nabla c) \cdot \bar{\mathbf{n}}]_{\Gamma_{sl}}=\left((D \nabla c)_s-(D \nabla c)_l\right) \cdot \bar{\mathbf{n}} \quad \forall \boldsymbol{x} \in \Gamma_{sl} ,
	\label{cs3}
\end{equation}
where $V_{\Gamma_{sl}}$ is the normal velocity of the interface, $\bar{\mathbf{n}}$ is the normal vector of the interface pointing from the solid phase to the liquid phase. It should be noted that the solute at the interface will be redistributed, satisfying $c_s=k_pc_l $, with $k_p $ being the partition coefficient, and thus we have 
\begin{equation}
	\left(1-k_p\right) c_{l, \Gamma_{sl}} V_{\Gamma_{sl}}=[(D \nabla c) \cdot \bar{\mathbf{n}}]_{\Gamma_{sl}}=\left(D \nabla c)_s-(D \nabla c)_l\right) \cdot \bar{\mathbf{n}} \quad \forall \boldsymbol{x} \in \Gamma_{sl} .
	\label{cs4}
\end{equation}
Actually, Eqs. (\ref{cs1})–(\ref{cs4}) constitute a sharp-interface model for the transport of solute concentration during the binary phase change process.

To avoid directly handling the boundary conditions on the solid-liquid phase change interface $\Gamma_{sl}$, we consider the concentration potential method proposed by Voller et al. \cite{VollerIJHMT2008}, where the concentration potential $V$ can be equivalent to a pseudo-concentration in the entire domain. This method simplifies the description of solute transport and interface conditions in the solid–liquid phase change of a binary fluid. By incorporating interfacial solute balance conditions implicitly into a convection-diffusion equation defined over a single domain, the method can significantly reduce numerical complexity and enhances computational efficiency, and has also been widely applied in the study of solid–liquid phase change problems \cite{VollerIJHMT2008-1, BhattacharyaJCP2014, SunIJHMT2016, JegatheesanCMS2021}. With this method, Eqs. (\ref{cs})-(\ref{cs4}) can be uniformly expressed as
\begin{equation}
	\frac{\partial c}{\partial t}+\nabla \cdot\left(c \mathbf{u}\right)=\nabla \cdot\left[D_e\left(\nabla V-\frac{(1-\phi) c}{\sqrt{2} \epsilon} \mathbf{n}\right)\right],
	\label{c}
\end{equation}
where $V = c / [ 1 - f_s (1 - k_p) ]$ is the pseudo-potential concentration, $c = f_s c_s + (1 - f_s) c_l$ is the total solute concentration, and $D_e = f_s k_p D_s + (1 - f_s) D_l$ is the effective diffusivity. Actually, Eq. (\ref{c}) can be considered as a unified framework for modeling solute transport during the solid-liquid phase transition of a binary fluid.

In summary, the present diffuse-interface model is composed of Eqs. (\ref{ACE})), (\ref{NS1}),(\ref{NS2}), (\ref{Tem}), and (\ref{c}), and can be used to describe the binary droplet freezing process. With the help of the order parameter $\phi$ and solid fraction $f_s$, the liquid, solid, and gas phases can be represented by ($\phi=1$ and $f_s=0$), ($\phi=1$ and $f_s=1$) and ($\phi=-1$ and $f_s=0$). In this case, the physical variables and parameters of the system can be characterized by a linear function of the order parameter and solid fraction,
\begin{equation}
	\zeta=\frac{1}{2} \phi\left(\zeta_l-\zeta_g\right)+f_s\left(\zeta_s-\zeta_l\right)+\frac{1}{2}\left(\zeta_l+\zeta_g\right)
\end{equation}
where the parameter $\zeta$ denotes the density, viscosity, thermal conductivity, and heat capacity.

\section{Numerical methods}
\label{sec3}
In this section, the LB method is used to solve the coupled governing equations for the flow field, phase field, temperature field, and concentration field. We refer the reader to Ref. \cite{ChaiPRE2020} for more details on the LB method for the Navier-Stokes and nonlinear convection-diffusion equations. For computational accuracy and efficiency of the LB method, we only adopt the Bhatnagar-Gross-Krook (BGK) and two-relaxation-time (TRT) models, which can also be extended to the more advanced multiple-relaxation-time (MRT) model \cite{ChaiPRE2020}.

\subsection{LB model for the phase field}
For the phase-field equation (\ref{ACE}), which can be considered as a convection-diffusion type equation with an extra flux term, the evolution of LB can be written as \cite{HuangPRE2024,HuangIJHMT2025}
\begin{equation}
	f_i\left(\mathbf{x}+\mathbf{c}_i \Delta t, t+\Delta t\right)-f_i(\mathbf{x}, t)=-\frac{1}{\tau_f}\left[f_i(\mathbf{x}, t)-f_i^{eq}(\mathbf{x}, t)\right]+\left(1-\frac{1}{2 \tau_f}\right) \Delta t F_i(\mathbf{x}, t),
\end{equation}
where $f_i(\mathbf{x}, t)$ is the order-parameter distribution function at position $\mathbf{x}$ and time $t$, $\mathbf{c}_i$ is the discrete velocity, $\Delta t$ is the time step. $\tau_f$ is the relaxation time related to the mobility $M$ through  $M=\eta c_s^2\left(\tau_f-0.5\right) \delta t$, $c_s = c /\sqrt 3$ is the lattice sound speed. The local equilibrium distribution function $f_i^{eq}$ and the discrete source term $F_i$ are given by
\begin{equation}
	f_i^{eq}=\omega_i \phi\left(1+\frac{\mathbf{c}_i \cdot \mathbf{u}}{c_s^2}\right), 	F_i=\frac{\omega_i \mathbf{c}_i \cdot\left[\partial_t(\phi \mathbf{u})+c_s^2 \frac{\nabla \phi}{|\nabla \phi|} \frac{1-\phi^2}{\sqrt{2} \epsilon}\right]}{c_s^2}+\omega_i \phi \nabla \cdot \mathbf{u}.
\end{equation}
The macroscopic order parameter in this model is calculated by
\begin{equation}
	\phi=\sum_i f_i+\frac{\Delta t}{2} \phi \nabla \cdot \mathbf{u}.
\end{equation}

\subsection{LB model for the flow field}

Unlike the original lattice BGK model for incompressible flows, we introduce an additional source term to express the effect of density change during the freezing process. In this case, the evolution equation of the LB method for the flow field reads \cite{HuangPRE2024}
\begin{equation}
	g_i\left(\mathbf{x}+\mathbf{c}_i \Delta t, t+\Delta t\right)-g_i(\mathbf{x}, t)=-\frac{1}{\tau_g}\left[g_i(\mathbf{x}, t)-g_i^{eq}(\mathbf{x}, t)\right]+\Delta t\left(1-\frac{1}{2 \tau_g}\right) G_i(\mathbf{x}, t),
\end{equation}
where $g_i(\mathbf{x}, t)$ is the distribution function for flow field, $\tau_g=\nu /c_s^2 \Delta t + 0.5$ is the relaxation time with $\nu=\mu / \rho$ being the kinematic viscosity. The local equilibrium distribution function $g_i^{eq}$ is defined by 
\begin{equation}
	g_i^{eq}= \begin{cases}\frac{p}{c_s^2}\left(\omega_i-1\right)+\rho s_i(\mathbf{u}), & \mathrm{i}=0 \\ \frac{p}{c_s^2} \omega_i+\rho s_i(\mathbf{u}), & \mathrm{i} \neq 0\end{cases}
\end{equation}
with
\begin{equation}
	s_i(\mathbf{u})=\omega_i\left[\frac{\mathbf{c}_i \cdot \mathbf{u}}{c_s^2}+\frac{\left(\mathbf{c}_i \cdot \mathbf{u}\right)^2}{2 c_s^4}-\frac{\mathbf{u} \cdot \mathbf{u}}{2 c_s^2}\right].
\end{equation}
In addition, to recover the governing equations (\ref{NS1}) and (\ref{NS2}) for fluid flows, the expression of $F_i$ should given by
\begin{equation}
	G_i=\omega_i\left[S+\frac{\mathbf{c}_i \cdot(\mathbf{F}+\rho \mathbf{f})}{c_s^2}+\frac{(\mathbf{u} \tilde{\mathbf{F}}+\tilde{\mathbf{F}} \mathbf{u}):\left(\mathbf{c}_i \mathbf{c}_i-c_s^2 \mathbf{I}\right)}{2 c_s^4}\right],
\end{equation}
where $S=\rho \dot{m}+\mathbf{u} \cdot \nabla \rho, \tilde{\mathbf{F}}=\mathbf{F}-\nabla p+c_s^2 \nabla \rho+c_s^2 \nabla \cdot S, \mathbf{F}=\mathbf{F}_{\mathrm{s}}+\mathbf{G}$ is the total force. The macroscopic density $\rho$, velocity $\mathbf{u}$ and pressure $p$ are computed by
\begin{equation}
	\rho \mathbf{u}^*=\sum \mathbf{c}_i g_i+\frac{\Delta t}{2} \mathbf{F}, \mathbf{u}=\mathbf{u}^*+\frac{\Delta t}{2} \mathbf{f},
\end{equation}
\begin{equation}
	p=\frac{c_s^2}{\left(1-\omega_0\right)}\left[\sum_{i \neq 0} g_i+\frac{\Delta t}{2} S+\tau \Delta t G_0+\rho s_0(\mathbf{u})\right],
\end{equation}
where $\mathbf{u}$ denotes the corrected velocity, $\mathbf{u}^*$ is the velocity without accounting for solid-liquid interactions, the force $\mathbf{f}=f_s \left(\mathbf{u}_s-\mathbf{u}^*\right) / \Delta t$ represents the solid-liquid interaction force with $\mathbf{u}_s=0$ being the solid-phase velocity \cite{LiuCICP2022, LiuCF2024}. This approach for solid-liquid interaction has been widely applied to investigate some problems, including the particulate flows \cite{LiuCICP2022, LiuCF2024}, solid-liquid phase change problems \cite{HuangPRE2024, HuangIJHMT2025} and dendrite growth \cite{ZhanCICP2023}.

\subsection{LB model for the temperature field}
To reduce numerical diffusion at the solid-liquid interface while maintaining a balance between computational accuracy and efficiency \cite{HuangJCP2015}, the TRT LB method is employed for the energy equation. The evolution equation of the LB method for the temperature field is given by \cite{LuIJTS2019}
\begin{equation}
	h_{i}\left(\mathbf{x}+\mathbf{e}_i \Delta t, t+\Delta t\right)=h_{i}(\mathbf{x}, t)-\frac{1}{\bar{\tau}_h}\left[\bar{h}_{i}(\mathbf{x}, t)-\bar{h}_{i}^{eq}(\mathbf{x}, t)\right]-\frac{1}{\tilde{\tau}_h}\left[\tilde{h}_{i}(\mathbf{x}, t)-\tilde{h}_{i}^{eq}(\mathbf{x}, t)\right]+\left(1-\frac{1}{2 \bar{\tau}_h}\right) \Delta t H_i(\mathbf{x}, t),
\end{equation}
where $\tilde{h}_i(\mathbf{x}, t)$ and $\bar{h}_i(\mathbf{x}, t)$ are the symmetric and anti-symmetric components of distribution function $h_i(\mathbf{x}, t)$ for the total enthalpy, $\tau_{\tilde{h}}$ and $\tau_{\bar{h}}$ are the corresponding relaxation times. Here the symmetric and anti-symmetric distribution functions are defined as \cite{LuIJTS2019}:
\begin{equation}
	\tilde{h}_i=\frac{h_i+h_{\bar{i}}}{2}, \bar{h}_i=\frac{h_i-h_{\bar{i}}}{2}, \tilde{h}_i^{e q}=\frac{h_i^{e q}+h_{\bar{i}}^{e q}}{2}, \bar{h}_i^{eq}=\frac{h_i^{e q}-h_{\bar{i}}^{e q}}{2},
\end{equation}
where $\bar{i}$ represents the opposite direction of $i$, and the opposite direction of $\mathbf{e}_0$ is itself. The equilibrium distribution function for the mixture total enthalpy $h_i^{eq}$ is given by
\begin{equation}
	h_i^{eq}= \begin{cases}\rho H-\rho C_{p, \text { ref }} T+\omega_i \rho C_p T\left(\frac{\rho_{\text { ref }} C_{p, \text { ref }}}{\rho C_p}-\frac{\mathbf{I}: \mathbf{u u}}{2 c_s^2}\right), & i=0, \\ \omega_i \rho C_p T\left[\frac{\rho_{\text { ref }} C_{p . \mathrm{ref}}}{\rho C_p}+\frac{\mathbf{e}_i \cdot \mathbf{u}}{c_s^2}+\frac{\left(\mathbf{e}_i \mathbf{e}_i-c_s^2 \mathbf{I}\right): \mathbf{u u}}{2 c_s^4}\right], & i \neq 0 .\end{cases}
\end{equation}
where $C_{p, \text { ref }}=2 C_{p, s} C_{p, l} /\left(C_{p, s}+C_{p, l}\right)$ is the reference specific heat. It should be noted that the anti-symmetric relaxation time is related to the thermal conductivity $\lambda/(\rho c_{p, \text {ref}})=\left(\bar{\tau}_h-0.5\right) c_s^2 \Delta t$, and satisfies the relation $1/\tilde{\tau}_h+1/\bar{\tau}_h=2 $ \cite{LuIJTS2019}. The additional source term $H_i(\mathbf{x}, t)$ is given by
\begin{equation}
	H_i(\mathbf{x}, t)=\rho C_p T \nabla \cdot \mathbf{u}.
\end{equation}
Finally, the mixture total enthalpy is determined by \cite{HuangPRE2024}
\begin{equation}
	\rho H=\sum_{i=0} h_i+\frac{\Delta t}{2}  \rho C_p T \nabla \cdot \mathbf{u}.
\end{equation}
After calculating the mixture total enthalpy, we can update the temperature and liquid fraction through Eq. (\ref{Hfl}).

\subsection{LB model for the concentration field}
For the transport of solute concentration during the freezing process, the evolution of the LB method with the BGK model can be written as \cite{ChaiPRE2020}
\begin{equation}
	l_i\left(\mathbf{x}+\mathbf{c}_i \Delta t, t+\Delta t\right)-l_i(\mathbf{x}, t)=-\frac{1}{\tau_l}\left[l_i(\mathbf{x}, t)-l_i^{eq}(\mathbf{x}, t)\right]+\Delta t\left(1-\frac{1}{2 \tau_l}\right) L_i(\mathbf{x}, t),
\end{equation}
where $l_i(\mathbf{x}, t)$ is the distribution function, and $l_i^{eq}(\mathbf{x}, t)$ is the equilibrium distribution function that can be  given by
\begin{equation}
	l_i^{eq}(\boldsymbol{x}, t)=\omega_i C \left(1+\frac{\mathbf{c}_i \cdot \mathbf{u}}{c_s^2}\right) .
\end{equation}
The discrete source term $L_i(\mathbf{x}, t)$ is expressed as
\begin{equation}
	L_i=\frac{\omega_i \mathbf{c}_i \cdot  [ \partial_t(C \mathbf{u})+  c_s^2 \mathbf{R}  ]     }{c_s^2},\quad \mathbf{R}=\frac{(1-\phi) C \mathbf{n} }{\sqrt{2} \epsilon} -\frac{(k_p-1) C \nabla(1-f_s)}{k_{\mathrm{eff}}},
\end{equation}
where $k_{\mathrm{eff}} = 1 - f_s (1 - k_p)$. The relaxation time $\tau_l$ is related to the effective diffusion coefficient  $D_e=c_s^2 \delta t\left(\tau_l-0.5\right) k_{\mathrm{eff}}$. In addition, the concentration is calculated by
\begin{equation}
	C=\sum_i l_i.
\end{equation}
It should be noted that the wettability boundary condition must be considered for problems involving three-phase contact lines. In this work, the implementation of the wettability condition follows the approach described in Ref. \cite{HuangPRE2024}, where the gradient and Laplacian terms are approximated by the following second-order isotropic difference schemes,
\begin{equation}
	\begin{aligned}
		\nabla \zeta(\mathbf{x}, t)=\sum_{i \neq 0} \frac{\omega_i \mathbf{c}_i \zeta\left(\mathbf{x}+\mathbf{c}_i \Delta t, t\right)}{c_s^2 \Delta t},\quad \nabla^2 \zeta(\mathbf{x}, t)=\sum_{i \neq 0} \frac{2 \omega_i\left[\zeta\left(\mathbf{x}+\mathbf{c}_i \Delta t, t\right)-\zeta(\mathbf{x}, t)\right]}{c_s^2 \Delta t^2} .
	\end{aligned}
\end{equation}

\section{Numerical results and discussion}
\label{sec4}

\subsection{Two-phase flows with Marangoni effect}

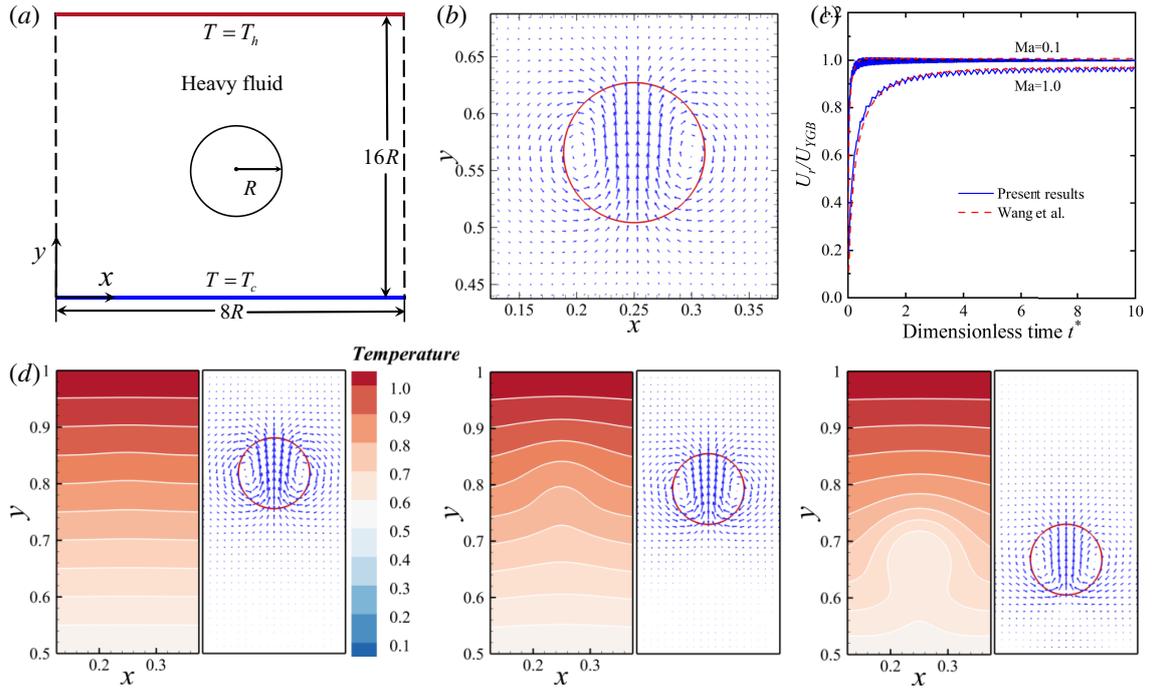
\begin{figure}[H]
	\centering
	\includegraphics[scale=0.4]{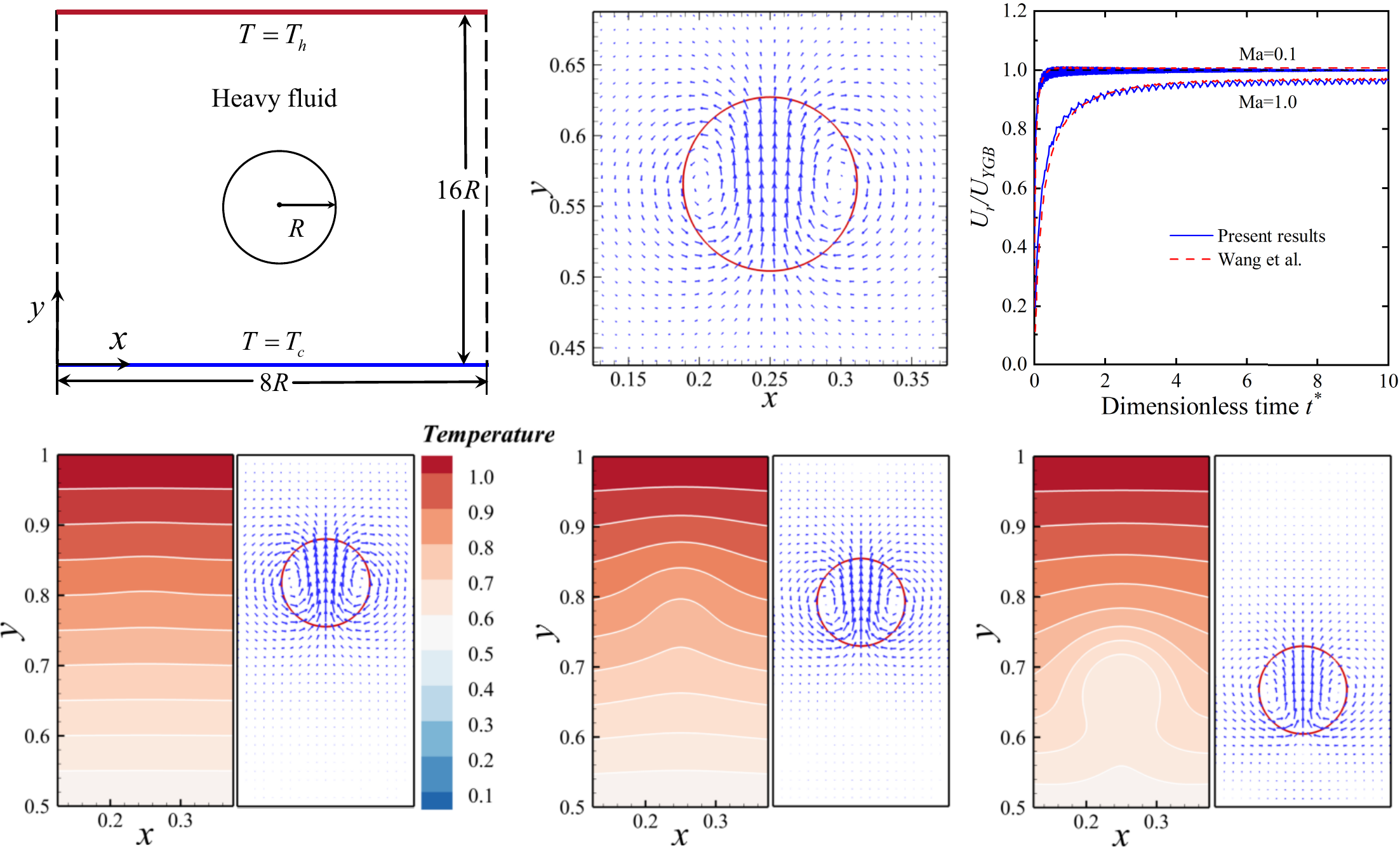} 
	\put(-425 ,250){(\textit{a})}
	\put(-265 ,250){(\textit{b})}
	\put(-125 ,250){(\textit{c})}	
	\put(-425 ,115){(\textit{d})}		
	\caption{(a) A schematic diagram of the droplet migration under the Marangoni effect. (b) The velocity around the rising droplet at $t^* = 10$. (c) A comparison of the normalized migration velocity dimensionless time $t^*$, the black dashed line indicates the theoretical prediction in the limit of vanishing Reynolds and Marangoni numbers. (d) The distributions of velocity and isotherm around the droplet at $Ma = 1$, $Ma = 10$ and $Ma = 100$, from left to right.}
	\label{fig2}
\end{figure}

In this part, we will test the diffuse-interface model and LB method through two benchmarks. We first consider the classical problem of droplet migration driven by a surface tension gradient, as analyzed by Young et al. \cite{YoungJFM1959} under infinitesimal Reynolds number and Marangoni number. For a droplet in an infinite medium with the equal thermal conductivities of the two phases, the steady-state migration velocity of the droplet can be expressed as 
\begin{equation}
	U_{YGB}=\frac{2U}{\left(2+3 \mu_l / \mu_h\right)\left(2+\alpha_l / \alpha_h\right)},
	\label{U_analy}
\end{equation}
where $U=-\sigma_T \nabla T_{\infty} R / \mu_h$ is the characteristic thermocapillary velocity, $R$ is the droplet radius and $\nabla T_{\infty}$ is the temperature gradient imposed on the system. The schematic of the problem is shown in Fig. \ref{fig2}(a), where a viscous droplet with the radius $R$ initially rests inside another heavy fluid, and is located at the center of an $ 8R \times 16R $ domain. A linear temperature field is imposed in the vertical direction with $ T_h = 32 $ and $ T_c = 0 $ (reference temperature $ T_{\text{ref}} $) on the top and bottom walls, resulting in a temperature gradient $ \nabla T_\infty = 0.1 $. For this problem, the periodic boundary conditions are applied on the left and right boundaries, while no-slip conditions are imposed on the top and bottom walls. The physical parameters are set as $ \sigma_{\text{0}} = 2.5 \times 10^{-3} $, $ \sigma_T = -10^{-4} $, $ \rho_l = \rho_h = 1 $, $ c_{p,l} = c_{p,h} = 1.0 $, $ \mu_l = \mu_h = 0.2 $, and $ \lambda_l = \lambda_h = 0.2 $. Besides, to describe the two-phase Marangoni flow, we introduce three dimensionless parameters, i.e., the Reynolds number $Re=LU/v_h $, Marangoni number $ Ma=\rho_h c_{p,h}LU/\lambda_h $ and capillary number $ Ca=U \mu_h/\sigma_{\text {0}} $. With the above specified parameters, one can determine the theoretical steady-state migration velocity $ U_{YGB} = 1.333 \times 10^{-4} $, $ Re = 0.1 $, $ Ma = 0.1 $, and $ Ca = 0.08 $. 

We present velocity field around the droplet in Fig. \ref{fig2}(b), where the surface tension gradient drives fluid flow from region of low surface tension (top) to that of high surface tension (bottom), which is consistent with the previous \cite{YoungJFM1959}. This would induce internal circulation in the droplet, causing the migration in the positive y-direction. In addition, we also measure the droplet velocity, 
\begin{equation}
	U_r(t)=\frac{\int_V \phi u_y d V}{\int_V \phi d V}=\frac{\sum_{\mathbf{x}} \phi(\mathbf{x}, t) u_y(\mathbf{x}, t)}{\sum_{\mathbf{x}} \phi(\mathbf{x}, t)}, \quad \phi>0 .
\end{equation}
and show the evolution of numerically predicted migration velocity in Fig. \ref{fig2}(c), in which the time has been normalized by $ R/U $. From this figure, one can see that the present results agree very well with those of Wang et al. \cite{WangPRE2023} and eventually converge to the theoretical solution. In addition, we also present the distributions of streamline and isotherm around the droplet in Fig. \ref{fig2}(d) where $ Re = 1.0 $ and $ \lambda_l = \lambda_h = 0.1 $. Other parameters remain unchanged, except for $ \mu_l = \mu_h = 0.1 $, $ \sigma_{\text{0}} = 5 \times 10^{-3} $ and $ \sigma_T = -2.5 \times 10^{-4} $ \cite{WangPRE2023}. It is found that at $ Ma = 1.0 $, the isotherms are almost parallel, indicating that for this case, the heat transfer is mainly dominated by diffusion. However, as $ Ma $ increases, the isotherms around the droplet bend upwards, indicating that the convective transfer in the system is enhanced, which is in agreement with previous studies \cite{LiuPRE2013,WangPRE2023}.

\begin{figure}[H]
	\centering
	\includegraphics[scale=0.6]{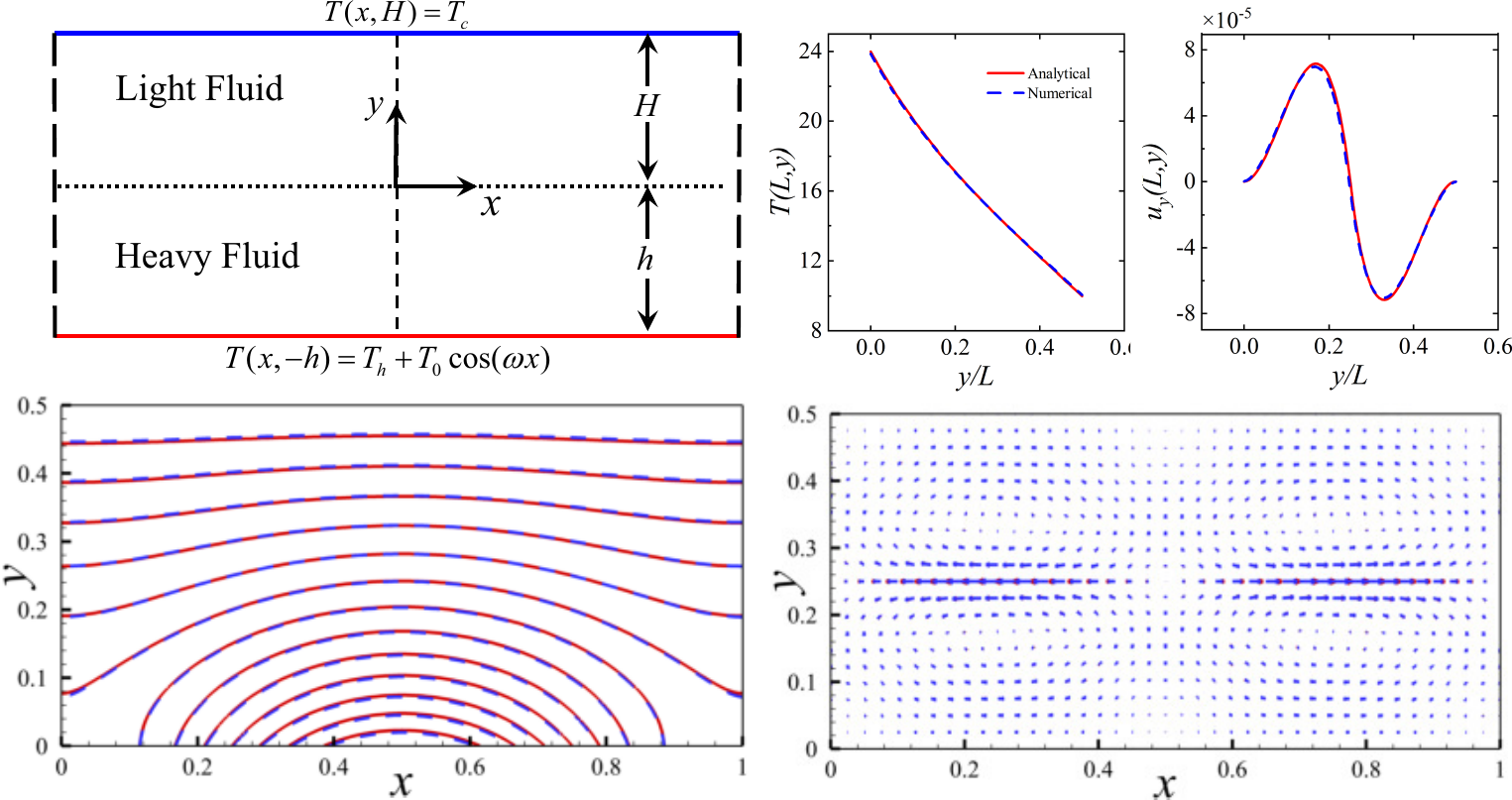} 
	\put(-450 ,220){(\textit{a})}
	\put(-226 ,220){(\textit{b})}	
	\put(-450 ,115){(\textit{c})}
	\put(-226 ,115){(\textit{d})}			
	\caption{ (a) A schematic diagram of the Marangoni convection of layered immiscible fluids. Velocity and temperature profiles along the center line (b). The distributions of temperature (c) and velocity (d) of thermocapillary flows with $\lambda_l / \lambda_h = 1.0$. The analytical and numerical results are indicated by red and blue, respectively. }
	\label{fig3}
\end{figure}
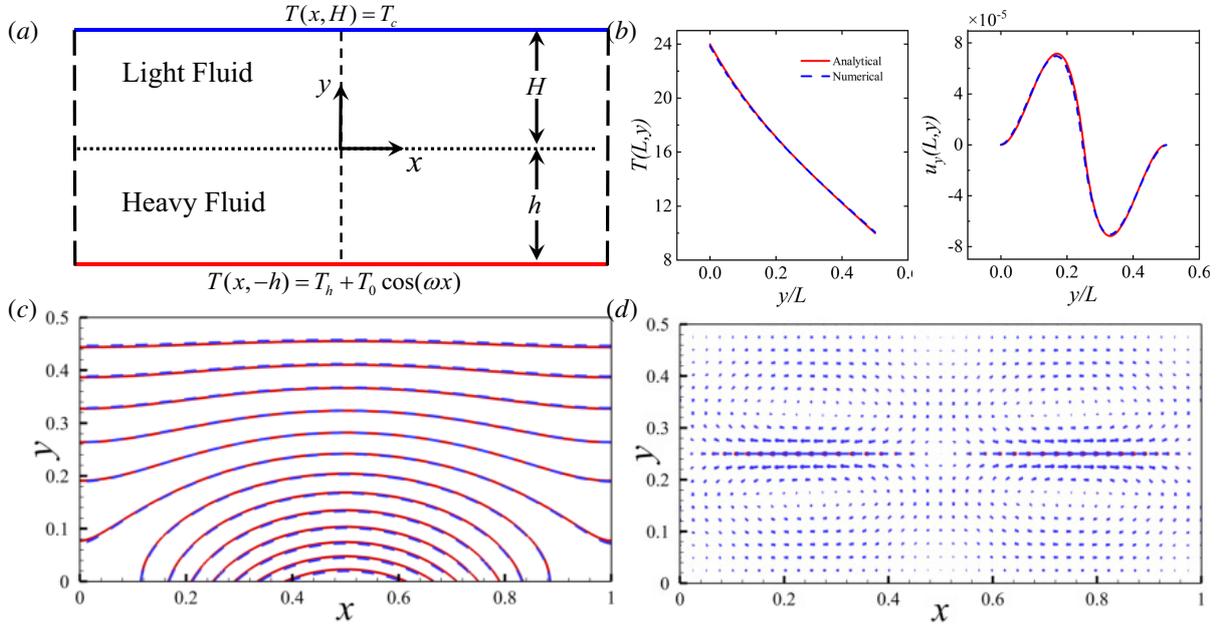

In the second problem, we consider the Marangoni convection in a vertically stratified system filled with two immiscible, light and heavy fluids, as illustrated in Fig. \ref{fig3}(a). The heights of the top and bottom parts are $H$ and $h$, respectively, while the problem is periodic in the x-direction. The temperatures at the top (\(y=H\)) and bottom boundaries \(y=-h\) are $T(x, H)=T_c$ and $T(x,-h)=T_h+T_0 \cos (2 \pi x / L)$, respectively, with $L$ being the channel length ($0<T_0<T_c<T_h$). In this case, the fluid typically moves from the hot region with low surface tension to the cold region with high surface tension, resulting in the Marangoni convection. For the case where convective transport of momentum and energy can be negligible and the interface remains flat (i.e., $Re \ll 1, Ma \ll 1$, and $Ca \ll 1$), the analytical solutions of the temperature and velocity can be obtained \cite{PendseIJTS2020}, and the details are presented in Appendix A.

We perform some simulations in a rectangular domain with $160\Delta x\times 80\Delta x$, $h=40 \Delta x$ and $H=40 \Delta x$. For the phase, temperature and flow fields, the bottom and top surfaces are the solid walls imposed by the no-flux, Dirichlet, and no-slip boundary conditions. The channel length $L$ is chosen as the characteristic length and the characteristic velocity is defined as $ U = \left|\sigma_T\right| T_0 h / (\mu_hL) $. The parameters used in the simulations are the same as those in some previous works \cite{LiuPRE2013,WangPRE2023}, i.e., $T_h=20, T_c=10, T_0=4, T_{\text {ref }}=10, \sigma_T=-5 \times 10^{-4}, \sigma_{\text {ref }}=2.5 \times 10^{-2}, \rho_l=\rho_h=1.0, c_{p,l}=c_{p,h}=1.0, \mu_l=\mu_h=0.2$ and $\lambda_h=0.2$. We plot the distributions of temperature and velocity in Figs. \ref{fig3} (c) and (d) where $\lambda_l / \lambda_h = 1.0$. As shown in these two figures, our numerical results (blue) agree well with the analytical solution (red) provided in Appendix A. We also present a quantitative comparison of numerical and analytical solutions in Fig. \ref{fig3}(b), which shows the velocity and temperature profiles along the center line. It is clear that there is a good agreement between them.

\subsection{Three-phase Stefan problem}

\begin{figure}[H]
	\centering
	\includegraphics[scale=0.45]{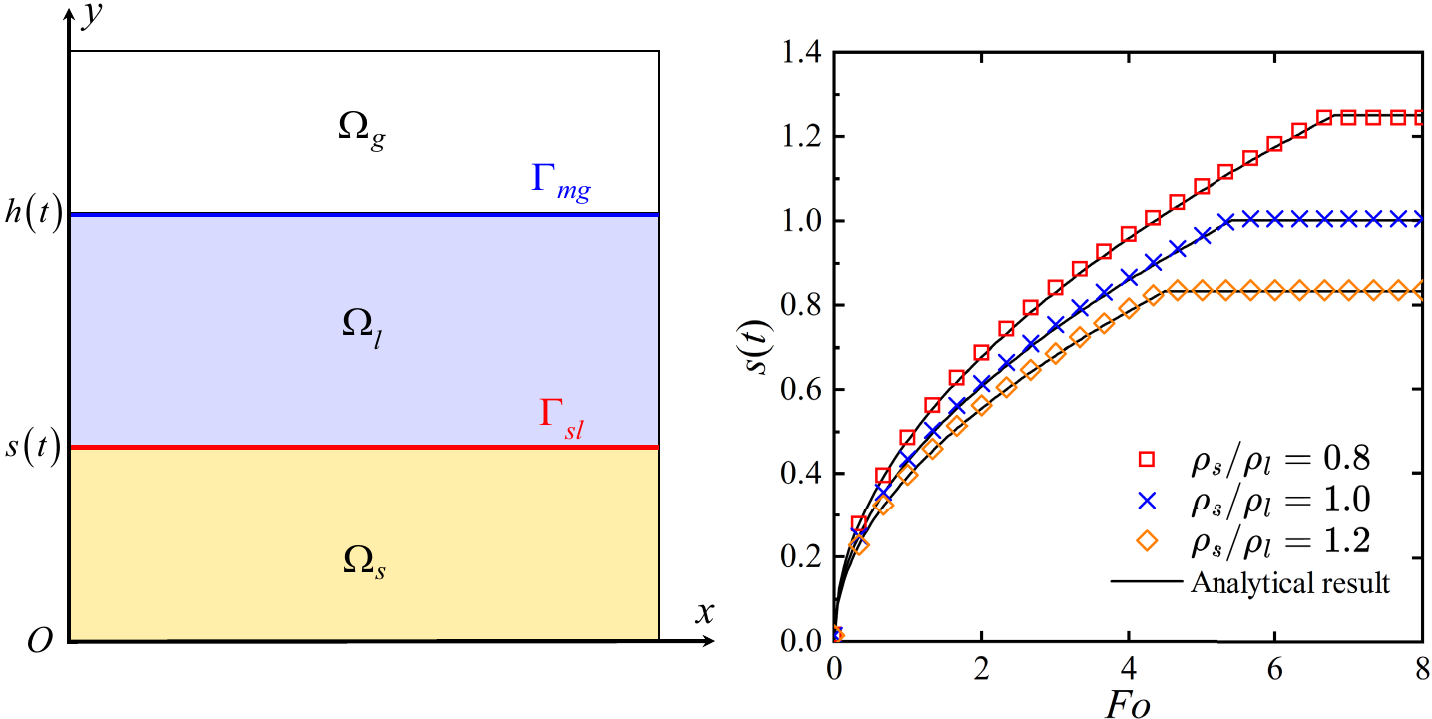} 
	\put(-350 ,160){(\textit{a})}
	\put(-175 ,160){(\textit{b})}		
	\caption{ (a) Schematic diagram of the three-phase Stefan problem, the domains occupied by gas, liquid and solid phases are denoted by $\Omega_g$ (white region), $\Omega_l$ (blue region), and $\Omega_s$ (yellow region), $\Gamma_{sl}$ indicates the freezing front, and $\Gamma_{mg}$ represents the interface between the phase-change material and gas phase. (b) Comparisons of the freezing front evolution between numerical and analytical solutions under different values of solid–liquid density ratio $\rho_s/\rho_l$ at $Ste = 0.1$ and $\alpha_r = 1.0$. }
	\label{fig4}
\end{figure}

We further test the accuracy of the present LB method for the phase-change volume expansion through simulating the solidification of a liquid film, as shown in Fig. \ref{fig4}(a). Initially, the cavity is filled with liquid, the remaining space is occupied by the gas phase, and the temperature is fixed at $T_h$. At $t > 0$, a cold temperature $T_w$ is imposed at the bottom wall ($y=0$), and the liquid phase begins to freeze into the solid phase. The periodic boundary condition is applied in the horizontal direction, while the bottom and top surfaces are the solid walls imposed by the no-flux, Dirichlet, and no-slip boundary conditions for the phase, temperature, and flow fields. According to the work of Lyu et al. \cite{LyuJCP2021}, the temporal evolution of the freezing front height $s(t)$ is $ s(t)=2 \kappa \sqrt{\alpha_s t}$, where $\kappa$ is the root of the following transcendental equation, 
\begin{equation}
	\frac{S t e}{\exp \left(\kappa^2\right) \operatorname{erf}(\kappa)}-\frac{\lambda_l\left(T_0-T_m\right) \sqrt{\alpha_r} \text { Ste }}{\lambda_s\left(T_m-T_w\right) \exp \left(\sqrt{\alpha_r} \kappa \rho_r\right)^2 \operatorname{erfc}\left(\sqrt{\alpha_r} \kappa \rho_r\right)}=\kappa \sqrt{\pi},
\end{equation}
where $\operatorname{erf}(x)=\frac{2}{\sqrt{\pi}} \int_0^x e^{-\eta^2} d \eta$ is the error function, $\operatorname{erfc}(x)=1-\operatorname{erf}(x)=$, $\rho_r=\rho_s / \rho_l$ and $\alpha_r=\alpha_s / \alpha_l$ are the ratios of the density and thermal diffusivity between the solid and liquid phases, respectively.

We first present some comparisons of freezing front between numerical results and analytical solutions for different solid-to-liquid density ratio $\rho_r = \rho_s / \rho_l$ in Fig. \ref{fig4}(b). It can be observed from this figure that the numerical results are in excellent agreement with the analytical solutions. Under the condition of $\rho_r=1$, the volume remains unchanged after solidification, the volume shrinkage takes place at $\rho_r>1$, while under the condition of $\rho_r<1$, the volume expansion occurs, which are consistent with the theoretical predictions. Furthermore, based on the mass conservation, the final height of the solid phase $h_f$ satisfies $h_f=\rho_l h_0 / \rho_s$. As shown in Table \ref{tab1}, the numerical results agree well with the theoretical values.

\begin{table}[H]
	\centering
	\caption{A comparison of the final solid height $h_f$ between the numerical and analytical data.}	
	\begin{tabular}{ccccclccc}
		\hline \hline
		\multirow{2}{*}{$\rho_l/\rho_s$} & \multirow{2}{*}{Analytical} & \multicolumn{3}{c}{Numerical}      &  & \multicolumn{3}{c}{Relative error}                                    \\ \cline{3-5} \cline{7-9} 
		&                             & $Ste=0.1$ & $Ste=0.15$ & $Ste=0.2$ &  & $Ste=0.1$             & $Ste=0.15$            & $Ste=0.2$             \\ \hline
		0.6                            & 1.667                       & 1.655     & 1.657      & 1.657     &  & $6.94 \times 10^{-3}$ & $5.56 \times 10^{-3}$ & $5.98 \times 10^{-3}$ \\
		0.7                            & 1.429                       & 1.425     & 1.425      & 1.424     &  & $2.59 \times 10^{-3}$ & $2.80 \times 10^{-3}$ & $3.01 \times 10^{-3}$ \\
		0.8                            & 1.250                       & 1.248     & 1.248      & 1.248     &  & $1.76 \times 10^{-3}$ & $1.84 \times 10^{-3}$ & $1.92 \times 10^{-3}$ \\
		0.9                            & 1.111                       & 1.111     & 1.111      & 1.111     &  & $2.70 \times 10^{-4}$ & $2.70 \times 10^{-4}$ & $2.70 \times 10^{-4}$ \\
		1.0                            & 1.000                       & 1.001     & 1.001      & 1.001     &  & $1.00 \times 10^{-3}$ & $1.00 \times 10^{-3}$ & $1.00 \times 10^{-3}$ \\
		1.1                            & 0.909                       & 0.913     & 0.913      & 0.912     &  & $4.40 \times 10^{-3}$ & $4.18 \times 10^{-3}$ & $3.19 \times 10^{-3}$ \\
		1.2                            & 0.833                       & 0.838     & 0.839      & 0.837     &  & $5.52 \times 10^{-3}$ & $6.60 \times 10^{-3}$ & $4.44 \times 10^{-3}$ \\ \hline \hline
	\end{tabular}
	\label{tab1}
\end{table}

\subsection{Stefan problem of binary solidification}

\begin{figure}[H]
	\centering
	\includegraphics[scale=0.4]{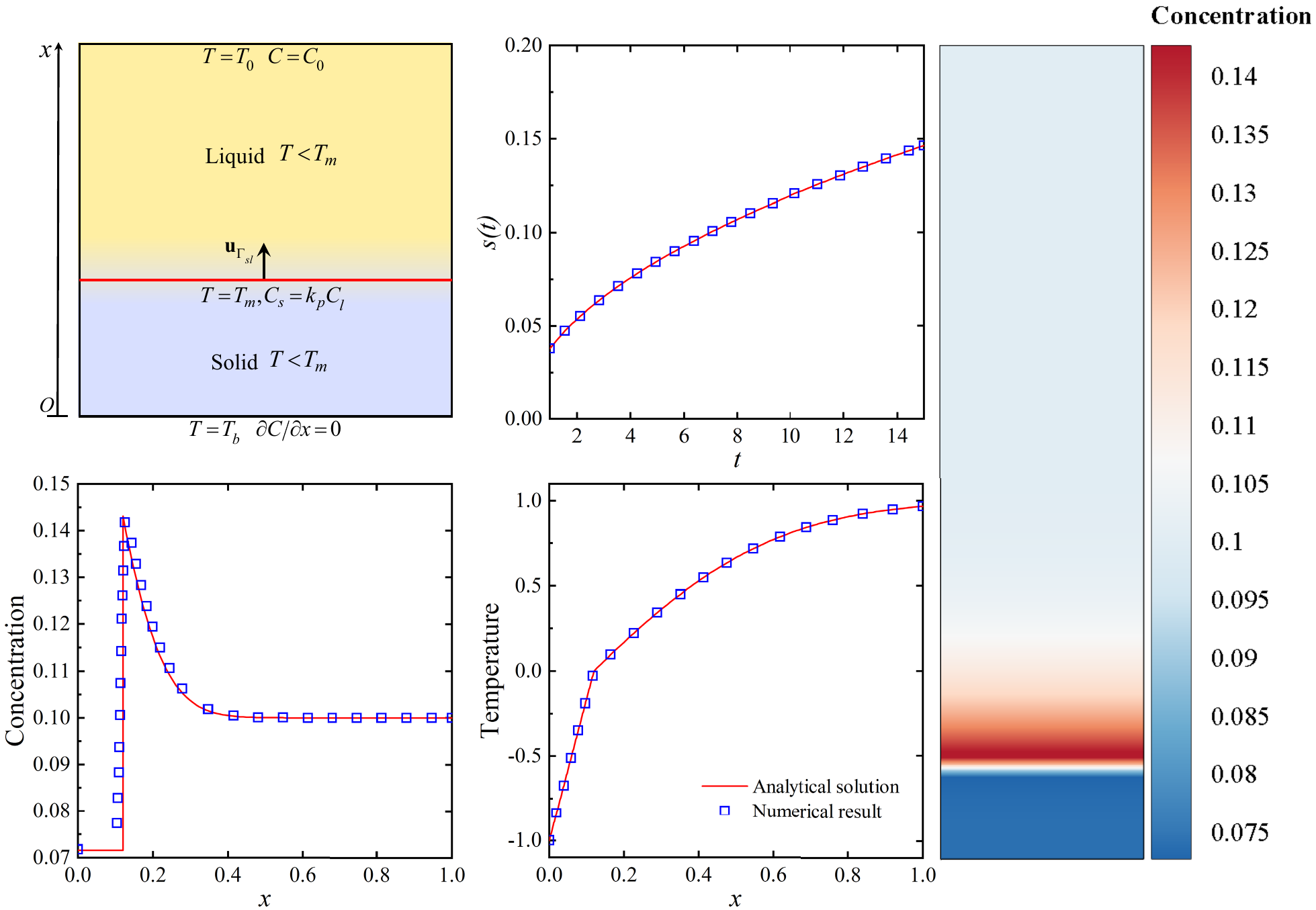} 
	\put(-380 ,240){(\textit{a})}
	\put(-243 ,240){(\textit{b})}	
	\put(-380 ,115){(\textit{c})}
	\put(-243 ,115){(\textit{d})}
	\put(-107 ,240){(\textit{e})}				
	\caption{ (a) A schematic diagram of binary solidification. (b) the evolution of the freezing front. (c) the distribution of solute concentration. (d) the distribution of temperature. (e) A macroscopic view of  concentration distribution.}
	\label{fig5}
\end{figure}
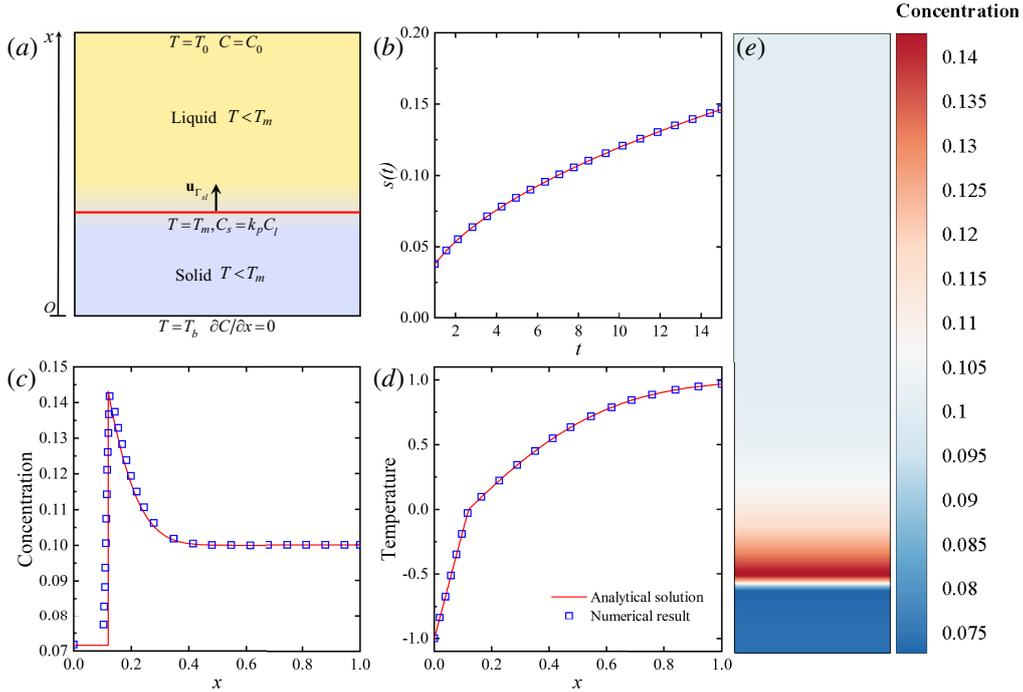

In this test case, we evaluate the accuracy of the present method in describing solute transport during the binary solidification. We consider a one-dimensional binary solidification problem in a semi-infinite domain $ x \geq 0 $, as shown in Fig. \ref{fig5} (a). Initially, the entire domain is full of the liquid phase with a uniform temperature $ T_0 $ and solute concentration $ C_0 $. At $ t = 0 $, the temperature at $ x = 0 $ is $ T_b < T_m $, where $ T_m $ is the melting temperature, and the concentration gradient at $ x = 0 $ is zero ($ \frac{\partial C}{\partial x} = 0 $). As $ x \to \infty $, the temperature and concentration remain $ T_0 $ and $ C_0 $, respectively. The solid-liquid interface is located at $ s(t) $, and the solid and liquid phases are in the regions $ 0 < x < s(t) $ and $ x > s(t) $. Under the above conditions, one can obtain the analytical solutions for the temperature and concentration \cite{AlexiadesBook2018}, 
\begin{equation}
	\begin{aligned}
		& T(x, t)= \begin{cases}T_b-\frac{T_b-T_{\mathrm{m}}}{\operatorname{erf} k} \operatorname{erf}\left(\frac{x}{2 \sqrt{\alpha_s t}}\right) & 0<x<s(t), t>0 \\
			T_0+\frac{T_{\mathrm{m}}-T_0}{\operatorname{erfc}\left(k \sqrt{\alpha_s / \alpha_l}\right)} \operatorname{erf}\left(\frac{x}{2 \sqrt{\alpha_l t}}\right) & x>s(t), t>0\end{cases}, \\
		& C(x, t)=\left\{\begin{array}{ll}
			\frac{k_p C_0 \exp \left(-k^2 \alpha_s / D_l\right)}{\exp \left(-k^2 \alpha_s / D_l\right)-\left(1-k_p\right) k \sqrt{\pi \alpha_s / D_l} \operatorname{erfc}\left(k \sqrt{\alpha_s / D_l}\right)} & 0<x<s(t), t>0 \\
			C_0+\frac{\left(1-k_p\right) C_0 k \sqrt{\pi \alpha_s / D_l}}{\exp \left(-k^2 \alpha_s / D_l\right)-\left(1-k_p\right) k \sqrt{\pi \alpha_s / D_l} \operatorname{erfc}\left(k \sqrt{\alpha_s / D_l}\right)} \operatorname{erfc}\left(\frac{x}{2 \sqrt{D_l t}}\right) & x>s(t), t>0
		\end{array},\right.
	\end{aligned}
\end{equation}
where $s(t)=2 k \sqrt{\alpha_s t}$ is the location of the solid-liquid interface, the parameter $k$ is the root of the following transcendental equation \cite{AlexiadesBook2018},
\begin{equation}
	\frac{C_{p, s}\left(T_m-T_{\mathrm{b}}\right)}{L \exp \left(k^2\right) \operatorname{erf}(k)}-\frac{C_{p, l}\left(T_0-T_m\right) \sqrt{\alpha_l / \alpha_s}}{L \exp \left(k^2 \alpha_s / \alpha_l\right) \operatorname{erfc}\left(k \sqrt{\alpha_s / \alpha_l}\right)}=k \sqrt{\pi} .
\end{equation}

In our simulations, the periodic boundary condition is imposed in the horizontal direction, and the parameters are set as $T_b=-1, T_m=0, T_0=0, C_{p,l}/C_{p,s}=1, L=10.0, C_0=0.1, k_p=0.5$. We present the evolution of the freezing front during the freezing process, as well as the distributions of temperature and concentration at $t=10$ in Fig. \ref{fig5}(b), (c) and (d). From these figures, one can see that the numerical results are close to the analytical solutions. Fig. \ref{fig5}(e) shows a macroscopic view of the concentration distribution at $t=10$, which clearly demonstrates that the solutes are rejected at the freezing front, leading to a significant increase of solute concentration in the liquid phase at the freezing front.

\subsection{Binary droplet freezing on a cold substrate}
In this part, the freezing process of the binary droplet on a cold substrate is numerically investigated. We first verify the accuracy of the present LB method in simulating pure droplet freezing on a cold substrate with different values of contact angle $\theta$, and perform a comparison with the previous work \cite{HuangN2025}. Initially, a droplet with the radius of $R$ is placed on the bottom substrate, and a lower temperature $T_w$ is imposed on the bottom boundary after the contact angle of the droplet is equal to the specified value $\theta$. The periodic boundary conditions are adopted for the left and right boundaries, and for the temperature field, the adiabatic boundary condition is applied on the top boundary, while for the flow field, the no-slip boundary conditions are imposed on both the top and bottom boundaries. The physical parameters are set to $\rho_s/\rho_l=0.9$, $C_{p,s}/C_{p,l}=1.0$ and $Ste=0.15$.

\begin{figure}[H]
	\centering
	\includegraphics[scale=0.4]{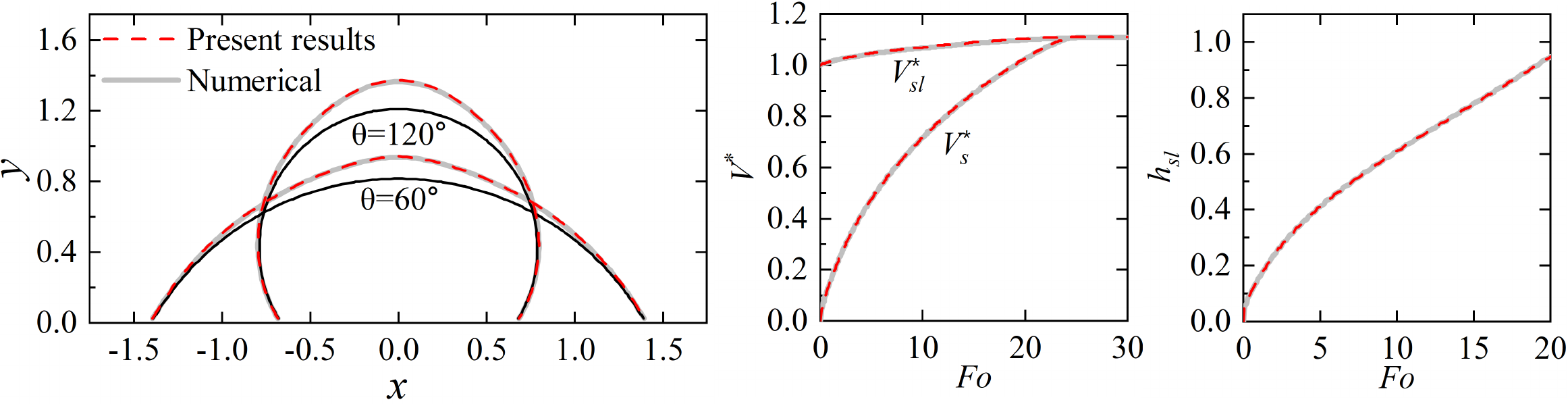} 
	\put(-450 ,110){(\textit{a})}
	\put(-242 ,110){(\textit{b})}		
	\put(-120 ,110){(\textit{c})}				
	\caption{ (a) Comparisons of present and previous numerical results of the final frozen droplet on a cold substrate with contact angles $\theta=60^\circ$ and $\theta=120^\circ$, the solid black lines represent the initial profiles of the droplet. (b) Evolution of dimensionless volume $V^*$, (c) Evolution of freezing front height $h_{sl}$ at $\theta=120^\circ$, $V^*_s$ and $V^*_{sl}$ denote the volumes of the solid phase and solid-liquid mixture, respectively. 
	}
	\label{fig6_1}
\end{figure}

We conduct some simulations, and plot the final contour of the frozen droplet, the volume evolutions of the solid phase and the solid-liquid mixture, and the height of the freezing front during the freezing process in Fig. \ref{fig6_1}. Due to the difference between solid and liquid densities ($\rho_s / \rho_l = 0.9 $), the volume of the solid-liquid mixture $V^*_{sl}$ expands during the freezing process. It is found that the final dimensionless volume is $ V^*_{sl} = 1.1107 $ and compared to the theoretical value, the error is about $ 3.7 \times 10^{-4} $. From the above discussion, one can find that present results are same as the previous works \cite{HuangPRE2024,HuangIJHMT2025}. The phenomenon of droplet volume expansion caused by the difference between the solid and liquid densities can be successfully captured, which also indicates that the present method is accurate in the study of pure droplet freezing on a cold solid surface.

\begin{figure}[H]
	\centering
	\includegraphics[scale=0.45]{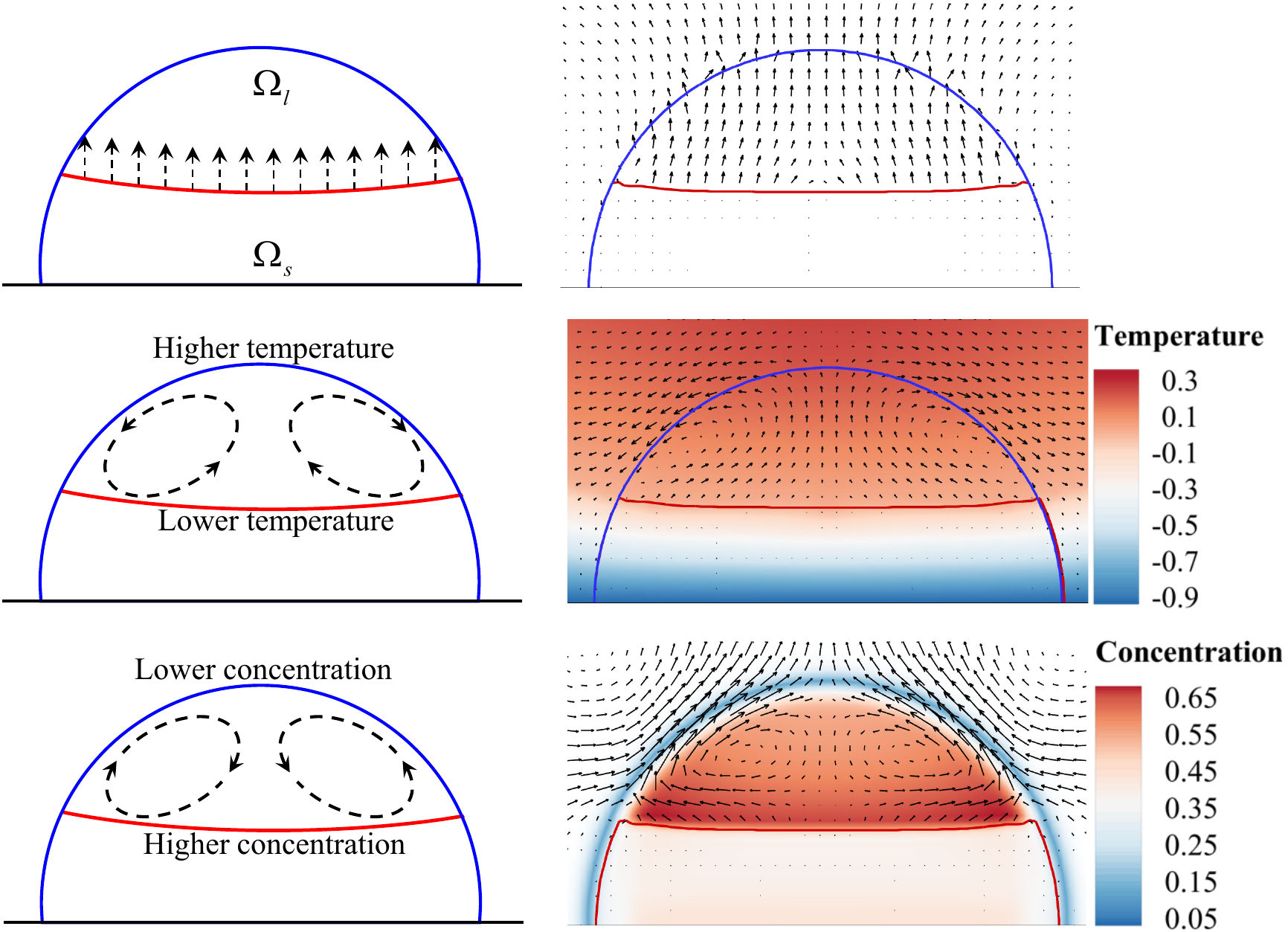} 
	\put(-380 ,260){(\textit{a})}
	\put(-380 ,170){(\textit{b})}
	\put(-380 ,75){(\textit{c})}
	\caption{ Comparison of flow pattern in a frozen droplet: (a) without Marangoni effect, (b) with thermally-driven Marangoni dominated effect and (c) solute-driven Marangoni dominated effect. Left: Schematic diagrams, right: Numerical results at $Fo = 0.42$. }
	\label{fig6_2}
\end{figure}

After that, we further investigate the freezing processes of binary droplet without the Marangoni flow, with thermally-driven Marangoni dominated flow, and solute-driven Marangoni dominated flow in \ref{fig6_2} where $ Fo = 0.42 $. In most of previous studies, the Marangoni effects on the droplet freezing are usually neglected, which results in a continuously rising freezing front with a weak upward flow both inside and outside the droplet, and there are no vortex structures formed in the flow field. However, when the Marangoni effect is considered, the internal flow patterns within the droplet undergo some significant changes, primarily influenced by the dominant Marangoni flow. For the case dominated by the thermally-driven Marangoni effect, the temperature in the region near the freezing front is lower, while it is much higher in the top or central region of the droplet, giving rise to a temperature gradient. Since the surface tension increases as temperature decreases, the surface tension near the freezing front (low-temperature region) is larger than that at the top region far from the front (high-temperature region). This surface tension gradient drives liquid flow from the high-temperature region with lower surface tension to the low-temperature region with higher surface tension, inducing a downward Marangoni flow (toward the freezing front) near the droplet interface. Conversely, for the case dominated by the solute-driven Marangoni effect, solute accumulates near the freezing front during the freezing process, leading to a higher solute concentration in the liquid phase near this region, and a lower concentration at the top region of droplet. Due to the fact that the surface tension decreases with the increase of solute concentration, the surface tension near the freezing front (high-concentration region) is lower than that at the top region of droplet (low-concentration region). Consequently, the surface tension gradient drives fluid flow from the high-concentration region with lower surface tension to the low-concentration region with higher surface tension, resulting in an upward solute-driven Marangoni flow (toward the top part the droplet) near the droplet interface. We note that these results are also qualitatively consistent with the experimental work \cite{WangPRL2024}.

\subsection{Binary solidification with impurity}

\begin{figure}[H]
	\centering
	\includegraphics[scale=0.45]{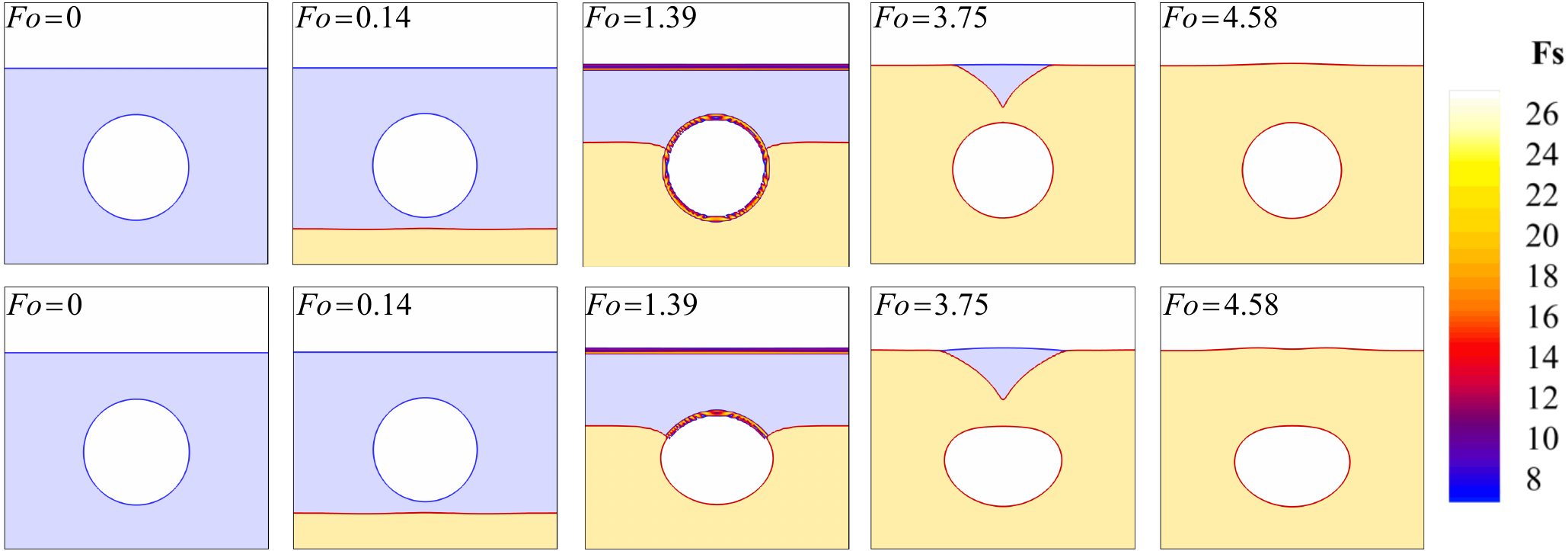} 
	\put(-460 ,150){(\textit{a})}
	\put(-460 ,70){(\textit{b})}					
	\caption{ The freezing process of liquid pool with a bubble, the gas phase, liquid phase, and solid phase are represented by white, blue, and yellow colors. (a) The traditional surface tension force $\mathbf{F}_s$, (b) the modified surface tension force $(1-f_s)\mathbf{F}_s$. The surface tension distribution at the phase interface is shown in $Fo=1.39$. }
	\label{fig7}
\end{figure}

The freezing process of fluids containing an insoluble impurity (e.g., some flexible bubbles, droplets, cells, and rigid particles) is a critical physical problem both in natural and industrial processes \cite{DuNRP2024,HuerreARFM2024}. The phase change dynamics of such a multiphase system is very complex, bringing some significant challenges for numerical methods. To assess the capacity of the present method in studying this type of freezing problem, we consider the freezing process in a liquid pool with a bubble. Within the phase-field framework, the surface tension at the gas-liquid interface is modeled by Eq. (\ref{Fs}), which can be derived from the energy-balance principle \cite{JacqminJCP1999} or the principle of least action \cite{YueJFM2004}. Huang et al. \cite{HuangJCP2022} pointed out that the traditional surface tension term acts simultaneously on the gas-solid-liquid three-phase interfaces. In order to eliminate its effect on the gas-solid interface, the surface tension is defined as $(1-f_s)\mathbf{F}_s$. Under the limitation of solid phase fraction $f_s$, the surface tension effect at the gas-solid interface is effectively suppressed, ensuring that the surface tension only acts on the gas-liquid interface. To give a comparison between the traditional and modified surface tension models, the freezing process of a bubble immersed in the liquid pool is investigated. As shown in Fig. \ref{fig7}(a), the computational domain is $L \times L$, the initial height of the liquid column is $H=3L/4$ and a bubble with the radius $0.2L$ is positioned at the center $(0.5L, 0.5L)$. The initial temperature is uniformly set to $T_0$, but the temperature at the bottom wall is fixed at a constant low temperature $T_w$. The physical boundary conditions are specified as follows: periodic boundary conditions are applied in the horizontal direction, while the no-slip and no-flux boundary conditions are imposed on the top and bottom boundaries. In our simulations, the half-way bounce-back scheme is used to treat the no-slip and no-flux boundary conditions, while the anti-bounce-back scheme is adopted to implement the Dirichlet boundary conditions of the temperature field.

We present the numerical results of the liquid pool freezing process under two different forms of surface tension ($\mathbf{F}_s$ and $(1-f_s)\mathbf{F}_s$) in Fig. \ref{fig7}, where the yellow, blue and white regions are filled with the solid, liquid and gas phases, respectively. At the initial stage, the solid phase starts to form near the cold substrate and the freezing front gradually moves upwards. As the solidification process develops, the bubble in the liquid pool is captured by the solid phase, and eventually a stable gas cavity is formed. It should be noted that the thermal conductivity of the gas phase is much smaller than that of the liquid phase, the solidification rate in the region above the captured bubble is relatively slowed down, resulting in the characteristic V-shaped structure at the moment $F_o=3.75$. In addition, a comparison of Fig. \ref{fig7}(a) and \ref{fig7}(b) also indicates that the modified surface tension force $(1-f_s)\mathbf{F}_s$ only acts at the gas-liquid interface (see the results at $Fo=1.39$), resulting in significant deformation of the bubbles during solidification (see the results at $Fo=3.75$). In contrast, the traditional surface tension force $\mathbf{F}_s$ acts on all interfaces (including the gas-liquid and solid-liquid interfaces, see the results at $Fo=1.39$).The force balance  effectively maintains the bubble’s equilibrium, allowing it to retain a nearly circular shape during the solidification process. These numerical results are qualitatively consistent with those reported in Ref. \cite{HuangJCP2022}.

\begin{figure}[H]
	\centering
	\includegraphics[scale=0.42]{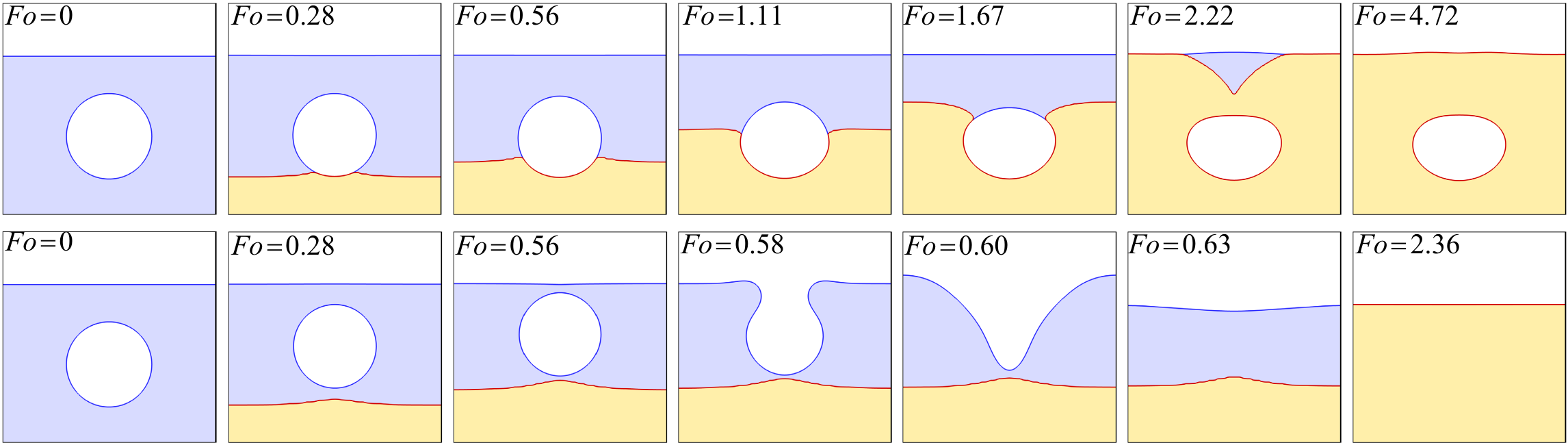} 
	\put(-478 ,125){(\textit{a})}
	\put(-478 ,55){(\textit{b})}					
	\caption{ The freezing process of a liquid pool with a bubble. (a) The case without Marangoni force and (b) the case with Marangoni force. The gas phase, liquid phase, and solid phase are denoted by white, blue, and yellow colors, respectively.}
	\label{fig8}
\end{figure}

\begin{figure}[H]
	\centering
	\includegraphics[scale=0.4]{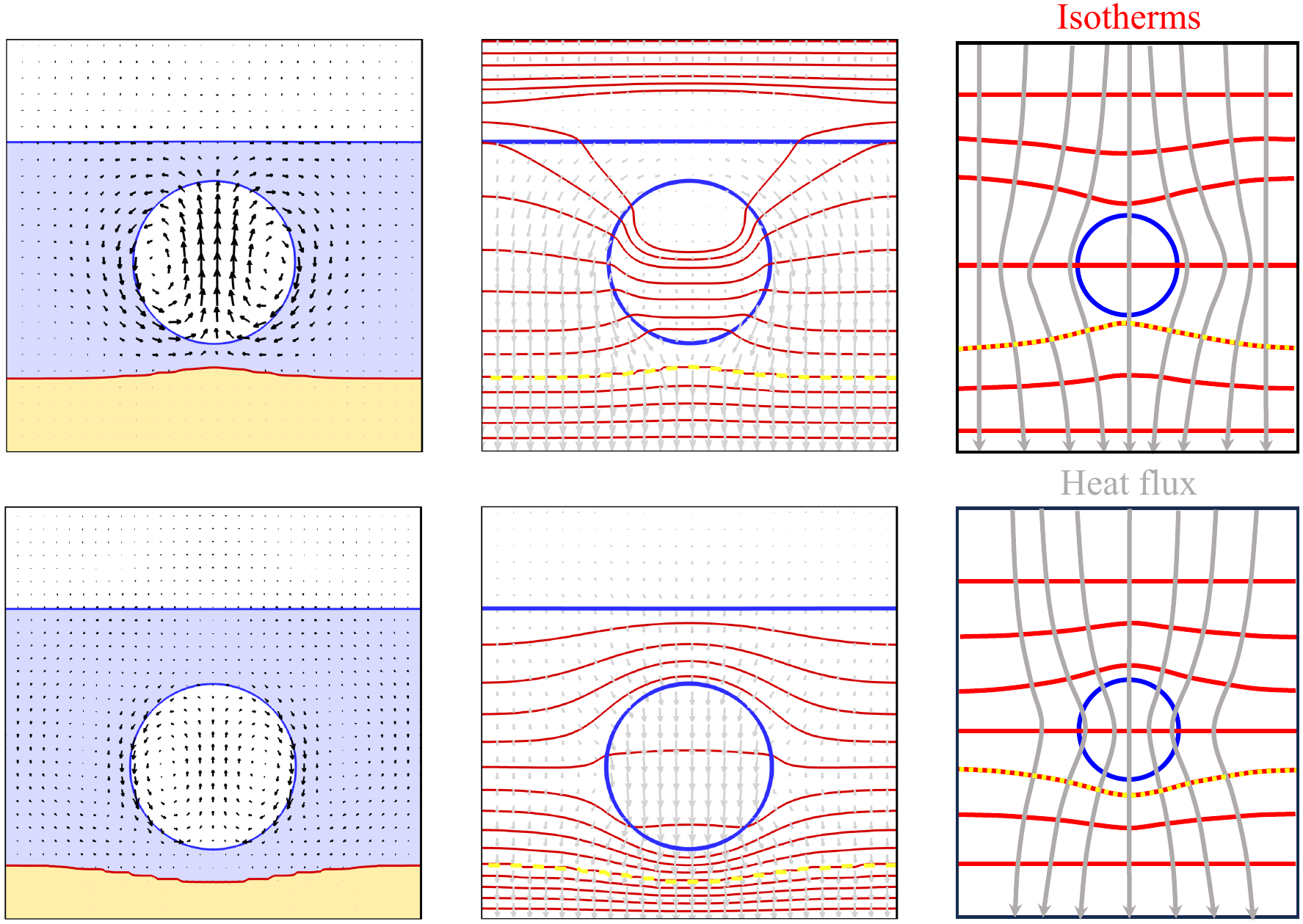} 
	\put(-355 ,225){(\textit{a})}
	\put(-355 ,102){(\textit{b})}					
	\caption{ Interaction between the freezing front and bubble under different values of thermal conductivity ratio ($\lambda_i/\lambda_l$): (a) $\lambda_i < \lambda_l$, $Fo=0.56$; (b) $\lambda_i > \lambda_l$, $Fo=0.21$. Left: flow field. Middle: isotherms (red solid line), heat flux distribution (gray arrow), and freezing front (yellow dashed line). Right: schematic of isotherm (red solid line), heat flux (gray solid line), and freezing front (yellow dashed line), adapted from Buuren et al. \cite{VanPRL2024} }
	\label{fig9}
\end{figure}

It should be noted that the role of Marangoni flow in the freezing process of an impurity-laden liquid pool has often been neglected in previous numerical studies \cite{ThirumalaisamyIJMF2023,ZhangJCP2024,HuangPRE2024,WeiJCP2025}. Here we also use this example to investigate the effect of Marangoni flow on the bubble migration and the evolution of the freezing front, and conduct a comparison with the case without including Marangoni force. As shown in Fig. \ref{fig8}, when Marangoni effect is considered, the bubble would continuously migrate upward and eventually escapes during the freezing process (see Fig. \ref{fig8}(b)). In contrast, when the Marangoni effect is not considered, the bubble is captured by the freezing front, forming a defect in solidification (see Fig. \ref{fig8}(a)). It is noteworthy that the freezing front exhibits some local protrusions near the bubble (see Fig. \ref{fig8}(b), $Fo = 0.28$ and $Fo = 0.56$), which is consistent with the results in some previous studies \cite{ParkJFM2006,VanPRL2024}. To further investigate the interaction between the freezing front and the bubble, we perform some simulations under different values of thermal conductivity ratio ($\lambda_i/\lambda_l$), and present the results in Fig. \ref{fig9} where flow and temperature fields as well as the distributions of heat flux are shown at $Fo=0.56$ and $Fo=0.21$. It is found that the shape of the freezing front is determined by the melting point isotherm of the liquid phase ($T=T_m$), while the distribution of isotherm is controlled by the thermal conductivity ratio. When the thermal conductivity of the bubble is lower than that of the surrounding liquid ($\lambda_i < \lambda_l$), the thermal resistance of the bubble is higher and the heat flow tends to bypass the bubble. This results in a bending of the isotherms near the bubble, causing the melting point isotherm (freezing front) to bulge toward the bubble.

On the contrary, when the thermal conductivity of the bubble is higher than that of the surrounding liquid ($\lambda_i > \lambda_l$), the heat flow tends to be conducted through the bubble with high thermal conductivity, resulting in bending of the isotherms away from the bubble. This is because the heat flow spreads out in the vicinity of the bubble, causing the melting point isotherm to move away from the bubble. In addition, it is observed that the distributions of the isotherms and heat flux in Fig. \ref{fig9} also agree well with those of Buuren et al. \cite{VanPRL2024}.

\section{Conclusions}
\label{sec5}
In this paper, a diffuse-interface model is proposed to describe the gas-liquid-solid multiphase flows involving solid-liquid phase change and solute transport, enabling quasi-equilibrium isotropic binary solidification without forming protrusions or dendritic crystal structures. In this model, the phase-field method and the enthalpy approach are employed to capture the gas-liquid interface, as well as the solid-liquid interface during the freezing process. Additionally, the solute transport between fluids is modeled by using a scalar concentration constraint approach, while the pseudo-potential concentration approach accurately captures solute redistribution at the solidification front. The model can strictly preserve mass conservation, defining the volume fractions of each phase through an order parameter and solid fraction, and account for volume change induced by the density difference during solid-liquid phase change. Furthermore, the Marangoni effect is also incorporated into the model, allowing it to rigorously degenerate into the classical phase-field model for incompressible two-phase flow and enthalpy model for binary solidification. To solve this diffuse-interface model, the LB method is developed, and validated through some benchmark problems, including the thermocapillary migration of a deformable droplet, the Marangoni convection in a stratified system, the three-phase Stefan problem and the binary liquid freezing problem. The method is further applied to study the freezing of a binary droplet on a cold wall, and it is found that the internal Marangoni flow patterns qualitatively agree with the previous works. Finally, the freezing process of a bubble-laden liquid pool is investigated, revealing that Marangoni flow has a significant effect on the bubble transport, and drives the bubble away from the freezing front. In particular, under the condition $\lambda_i < \lambda_l$, the freezing front protrudes toward the bubble, while under the condition $\lambda_i > \lambda_l$, it deflects away from the bubble. These results are also consistent with those reported in the experimental studies.

\section*{Appendix A. An analytical solution of the Marangoni convection in a vertically stratified system }
\label{ApA}

In this appendix, we present the analytical solutions of the temperature and velocity of thermocapillary-driven (Marangoni) convection in a stratified system \cite{PendseIJTS2020}. Under the condition of
creeping flow and negligible thermal convection, the solutions of temperature and velocity of the upper fluid are 
\begin{subequations}
	\begin{equation}
		T^L(x, y)=\frac{\left(T_c-T_h\right) y+\lambda_r T_c h+T_h H}{H+\lambda_r h}+T_0 f\left(\tilde{H}, \tilde{h}, \lambda_r\right) \sinh (\tilde{H}-\omega y) \cos (\omega x),
	\end{equation}
	\begin{equation}
		u_x^L(x, y)=U_{\max }\left\{\left[C_1^H+\omega\left(C_2^H+C_3^H y\right)\right] \cosh (\omega y)+\left(C_3^H+\omega C_1^H y\right) \sinh (\omega y)\right\} \sin (\omega x),
	\end{equation}
	\begin{equation}
		u_y^L(x, y)=-\omega U_{\max }\left[C_1^H y \cosh (\omega y)+\left(C_2^H+C_3^H y\right) \sinh (\omega y)\right] \cos (\omega x),
	\end{equation}\textbf{}	
\end{subequations}
and in the lower fluid is
\begin{subequations}
	\begin{equation}
		T^H(x, y)=\frac{\lambda_r\left(T_c-T_h\right) y+\lambda_r T_c h+T_h H}{H+\lambda_r h}+T_0 f\left(\tilde{H}, \tilde{h}, \lambda_r\right)\left[\sinh (\tilde{H}) \cosh (\omega y)-\lambda_r \sinh (\omega y) \cosh (\tilde{H})\right] \cos (\omega x),
	\end{equation}
	\begin{equation}
		u_x^H(x, y)=U_{\max }\left\{\left[C_1^h+\omega\left(C_2^h+C_3^h y\right)\right] \cosh (\omega y)+\left(C_3^h+\omega C_1^h y\right) \sinh (\omega y)\right\} \sin (\omega x),
	\end{equation}
	\begin{equation}
		u_y^H(x, y)=-\omega U_{\max }\left[C_1^h y \cosh (\omega y)+\left(C_2^h+C_3^h y\right) \sinh (\omega y)\right] \cos (\omega x),
	\end{equation}
\end{subequations}
where some parameters appeared in the above equations are defined as
\begin{equation}
	\tilde{H}=H \omega, \quad \tilde{h}=h \omega, \quad f\left(\tilde{H}, \tilde{h}, \lambda_r\right)=\left[\lambda_r \sinh (\tilde{h}) \cosh (\tilde{H})+\sinh (\tilde{H}) \cosh (\tilde{h})\right]^{-1},
\end{equation}

\begin{equation}
	\begin{aligned}
		C_1^H=\frac{\sinh ^2(\tilde{H})}{\sinh ^2(\tilde{H})-\tilde{H}^2}, \quad C_2^H=\frac{-H \tilde{H}}{\sinh ^2(\tilde{H})-\tilde{H}^2}, \quad C_3^H=\frac{2 \tilde{H}-\sinh (2 \tilde{H})}{2\left[\sinh ^2(\tilde{H})-\tilde{H}^2\right]}, \\
		C_1^h=\frac{\sinh ^2(\tilde{h})}{\sinh ^2(\tilde{h})-\tilde{h}^2}, \quad C_2^h=\frac{-h \tilde{h}}{\sinh ^2(\tilde{h})-\tilde{h}^2}, \quad C_3^h=\frac{\sinh (2 \tilde{h})-2 \tilde{h}}{2\left[\sinh ^2(\tilde{h})-\tilde{h}^2\right]},
	\end{aligned}
\end{equation}
and
\begin{equation}
	U_{\max }=-\left(\frac{T_0 \sigma_T}{\mu_h}\right) g\left(\tilde{H}, \tilde{h}, \lambda_r\right) h\left(\tilde{H}, \tilde{h}, \lambda_r\right),
\end{equation}
where
\begin{equation}
	g\left(\tilde{H}, \tilde{h}, \lambda_r\right)=\sinh (\tilde{H}) f\left(\tilde{H}, \tilde{h}, \lambda_r\right),
\end{equation}
and
\begin{equation}
	h\left(\tilde{H}, \tilde{h}, \mu_r\right)=\frac{\left[\sinh ^2(\tilde{H})-\tilde{H}^2\right]\left[\sinh ^2(\tilde{h})-\tilde{h}^2\right]}{\mu_r\left[\sinh ^2(\tilde{h})-\tilde{h}^2\right][\sinh (2 \tilde{H})-2 \tilde{H}]+\left[\sinh ^2(\tilde{H})-\tilde{H}^2\right][\sinh (2 \tilde{h})-2 \tilde{h}]} .
\end{equation}

\section*{Acknowledgments}
This research was supported by the National Natural Science Foundation of China (Grants No. 123B2018 and No. 12501599), the Interdisciplinary Research Program of HUST (2024JCYJ001 and 2023JCYJ002), and the Fundamental Research Funds for the Central Universities, HUST (No. 2024JYCXJJ016). The computation was completed on the HPC Platform of Huazhong University of Science and Technology.


\end{document}